\documentclass[aps,reprint,groupedaddress]{revtex4-1}
\usepackage{graphicx, amsmath, epstopdf, bm, amsthm, amssymb}% Include figure files
\graphicspath{{./images/}}

\begin{document}

%\preprint{}

\title{A globally stable attractor that is locally unstable everywhere}



\author{Phanindra Tallapragada}%
\email{ptallap@clemson.edu}
\affiliation{Department of Mechanical Engineering, Clemson University, Clemson, SC, U.SA, 29607}

\author{Senbagaraman Sudarsanam}
%\email{ssudars@g.clemson.edu}
\affiliation{Department of Mechanical Engineering, Clemson University, Clemson, SC, U.S.A, 29607}


\date{\today}

\begin{abstract}
We construct two examples of invariant manifolds that despite being locally unstable at every point in the transverse direction are globally stable. Using numerical simulations we show that these invariant manifolds temporarily repel nearby trajectories but act as global attractors.  We formulate an explanation for such global stability in terms of the `rate of rotation' of the stable and unstable eigenvectors spanning the normal subspace associated with each point of the invariant manifold. We discuss the role of this rate of rotation on the transitions between the stable and unstable regimes.

\end{abstract}

%\keywords{}

\maketitle
%\section{Introduction}
\section{\label{sec:level2}Introduction}
Invariant manifolds organize  the trajectories of a dynamical system and determine in a qualitative manner the behavior of a system. % For example the fixed points of a dynamical system can determine the existence of periodic behavior or if  the state of the system converges to or diverges from an equilibrium state. Higher dimensional invariant manifolds such as limit cycles in a plane can also determine whether a dynamical system admits periodic behavior.
An invariant manifolds can be thought of as unstable if nearby trajectories diverge away from it and stable otherwise. The global stability of an invariant manifold in $\mathbb{R}^n$, $n \geq 3$, can be assessed using Lyapunov type numbers that were first described by Fenichel, \cite{Fenichel_1971}. However a globally stable invariant  manifold can still have regions of instability, where nearby trajectories temporarily diverge from the invariant manifold. To characterize this local instability Haller proposed the use of normal infinitesimal Lyapunov  exponent (NILE), \cite{Haller_siam_2010}, which are similar to local Lyapunov numbers, \cite{abarbanel_jnls_1992}. Subsets of the invariant manifold where the NILE is positive are locally unstable, with nearby trajectories jumping away from the invariant manifold.

We construct two examples of dynamical systems with invariant manifolds that despite being unstable at every point in the normal direction are nevertheless globally stable.  The examples are inspired by some recent findings, \cite{sspt_cnsns_2017}, on the dynamics of inertial particles modeled using a simplified Maxey-Riley equation, \cite{maxey-riley, babiano, Tallapragada2008}. The planar motion of inertial particles in a fluid generates a four dimensional dynamical system with the fluid streamlines forming the invariant manifold for a time independent fluid flow. Though this invariant manifold is globally stable, it was observed that this globally attracting invariant manifold contained sub-domains of local instability \cite{babiano,haller_pof_2008}. In \cite{sspt_cnsns_2017} it was hypothesized that locally unstable subsets of the invariant manifold are globally stable due to the `rotation' of the stable and unstable eigenvectors of an associated reduced two dimensional system along a particles trajectory. The physical origin of this rotation of the stable and unstable eigenvectors lies in the vorticity field of the fluid. 

The two dynamical systems we discuss in this paper  are constructed by explicitly using this phenomenon of the rotation of the stable and unstable eigenvectors that span the normal subspace associated with each point of the invariant manifold.  We also demonstrate  via numerical simulations the relationship between the global stability of the invariant manifold and the rate of rotation of these eigenvectors. The first system has an invariant manifold that is not compact, while the second system has a limit cycle. In both cases the invariant manifolds repel a subset (of nonzero measure) of  trajectories in any neighborhood for a brief period of time. However every trajectory eventually converges to the invariant manifold. The concrete examples in this paper demonstrate a novel type of a global attractor that is locally unstable everywhere. 

It is important to draw attention to past work that demonstrate the existence  of limit cycles that are locally unstable but globally (or orbitally) stable. For instance \cite{kurrer_schulten_1991, ali_menzinger_chaos99} discuss dynamical systems with a limit cycle a subset of which is locally unstable and a subset of which is locally stable. Some perturbations normal to a subset of the limit cycle would grow, while in the stable subset of the limit cycle, all perturbations would decay. The limit cycle is however globally or orbitally stable. Hybrid dynamical systems can also demonstrate such limit cycles. For instance \cite{ross_physicaD_2008} shows that the hybrid dynamical system of the so called passive walker has a locally unstable limit cycle, but one that is globally stable due to a reset map that makes the dynamical system piecewise smooth. In contrast in this paper we construct smooth dynamical systems with invariant sets that are locally unstable everywhere and yet are globally stable. More importantly, the constructive examples in this paper distill the mechanism by which an invariant set that is locally unstable could become globally stable. 

The paper is organized as follows. In section \ref{sec:nile} we review the definition of the normal stability of an invariant manifold. This stability is in a local sense and is measured by the  normal infinitesimal Lyapunov exponent (NILE),  \cite{Haller_siam_2010}. This indicator of stability has also been referred to as the local Lyapunov exponent, \cite{abarbanel_jnls_1992, ali_menzinger_chaos99}. In \ref{sec:ex1} we construct a dynamical system in $\mathbb{R}^3$ that has a non compact invariant manifold that is locally unstable everywhere but is globally stable. In \ref{sec:limit_cycle} we construct a dynamical system in $\mathbb{R}^3$ that has a limit cycle that is locally unstable everywhere but is globally stable. 

\section{Normal local stability of an invariant manifold} \label{sec:nile}
Consider a time invariant dynamical system of the form
\begin{equation}\label{eq:sys}
\dot{\xi} = f(\xi), \;\;\;\;\;\;  \xi \in \mathbb{R}^n
\end{equation}
where $n \geq 2$ and let a trajectory of the system, $\xi(t; t_0, \xi_0)$ be defined as the solution to the initial value problem, $\xi(t_0) = \xi_0$. The flow map,  $\phi_{t_0}^t  : \mathbb{R}^n \mapsto \mathbb{R}^n$, associated with this system is defined as the transformation
\begin{equation} \label{eq:flow_map}
\phi_{t_0}^t (\xi_0) = \xi(t; t_0, \xi_0) .
\end{equation} 
A $m$ dimensional submanifold $M\ \subset \mathbb{R}^n$ is defined as invariant under the flow if 
\begin{equation}
\phi^t_{t_0}(M) \subset M.
\end{equation}
The manifold $M$ is a minimal invariant manifold if no proper subset of $M$ is invariant.

Let $T_{\xi}M$ and $N_{\xi}M$ denote the tangent and normal spaces to $M$ at $\xi \in M$, i.e., $T_{\xi}\mathbb{R}^n = T_{\xi}M \oplus N_{\xi}M$.  We will denote the projection from the tangent space of $\mathbb{R}^n$ at $\xi$ to $N_{\xi}M$ by 

\begin{align} \label{eq:projection}
	\Pi ^{t}_{\xi}:T_{\xi}\mathbb{R}^n =T_{\xi}\mathcal{M}\oplus N_{\xi}\mathcal{M} &\rightarrow N_{\xi}\mathcal{M}.\\ 
(u_t,u_n) &\mapsto u_n \nonumber
\end{align}

Denoting the  Jacobian of the flow map ${\phi}_t^{t+s}$ by $D{\phi}_t^{t+s}$, the evolution of normal vectors along a trajectory is given by
\begin{equation}
\Pi ^{t+s}_{F^{t+s}_t(\xi)} D{\phi}_t^{t+s}  u_n(t)=  u_n(t+s)
\end{equation} 
where  $u_n(t) \in N_{\xi(t)}M$ and $u_n(t+s) \in N_{\xi(t+s)}M$. The instantaneous growth in the norm of normal vectors is captured by the so called  normal infinitesimal Lyapunov exponent (NILE),  defined as, \cite{Haller_siam_2010},
\begin{equation}
	\sigma(\xi,t) = \lim_{s \to 0^+} \frac{1}{s} \log \| \Pi ^{t+s}_{{\phi}^{t+s}_t(\xi)} D{\phi}_t^{t+s}|_{N_{\xi} \mathcal{M}} \|.
\end{equation}
The normally stable and unstable sets $M_s \in M$ and $M_u \subset M$ respectively are defined as
\begin{align}\label{eq:set_defn}
M_s &= \{\xi \in M | \sigma(\xi,t) <0\} \nonumber \\ 
M_u &= \{\xi \in M | \sigma(\xi,t) >0\} .
\end{align}

\begin{figure}[!h]
	\begin{center}
		\includegraphics[width=0.65\hsize]{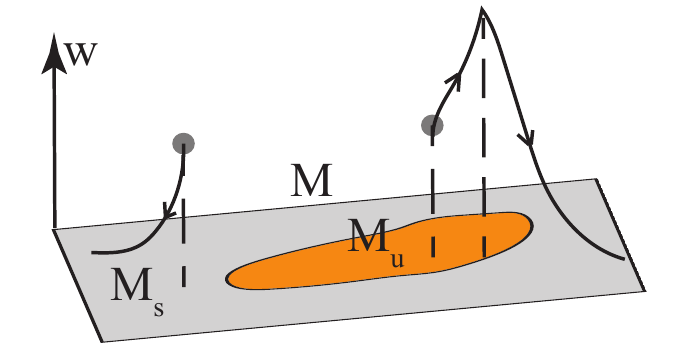}
	\end{center}
	\caption{Normal perturbations from the stable subsets $M_s \subset M$ decay monotonically, while normal perturbations from the unstable subsets,$M_u \subset M$ increase. When the projection of a normally perturbed trajectory leaves $M_u$ and enters $M_s$, the perturbations decay.}\label{fig:Ms_Mu}
\end{figure}

The stable sets, $M_s$, are regions of $M$ where every normal perturbation contracts in norm as illustrated in fig.\ref{fig:Ms_Mu} while the unstable sets $M_u$ are regions of $M$ where at least a subset (of positive measure) of  normally perturbed trajectories diverge from $M$. The NILE, unlike the Lyapunov-like numbers of Fenichel, \cite{Fenichel_1971} characterizes the local stability of subsets of $M$.  The NILE has the advantage that it extracts information about the local stability of any point on an invariant manifold and is easy to compute numerically.

\section{A gobally stable non compact invariant manifold that is locally unstable everywhere} \label{sec:ex1}

Consider a vector field in  $\mathbb{R}^3$  of the form

\begin{align} \label{eq:diff_eq}
\begin{pmatrix}
\dot{x}\\
\dot{y}
\end{pmatrix} &= 
\textbf{A}(z)\begin{pmatrix}
x\\
y
\end{pmatrix} \nonumber \\
\dot{z} &= g(z)
\end{align}
where $\textbf{A}(z)$ is a two dimensional square matrix whose entries are dependent on $z$. The manifold 
\begin{equation}\label{eq:invariant_set}
M = \{(x,y,z) | x=0, y=0 \}
\end{equation}
which is a straight line (the `z- axis') is invariant. In the simple case where

\[
\textbf{A}(z) = \begin{pmatrix}
\lambda_1 & 0\\
0 & \lambda_2
\end{pmatrix}
\]
such that $\lambda_1<0$, $\lambda_2>0$ and $\lambda_1+\lambda_2<0$, all trajectories eventually diverge away from the $z$-axis, making the invariant manifold $M$ unstable.

We now modify the matrix $\textbf{A}(z)$ in the following way,
\begin{equation} \label{eq:A}
\textbf{A}(z) = \textbf{R}\begin{pmatrix}
\lambda_1 & 0\\
0 & \lambda_2
\end{pmatrix} \textbf{R}^{-1}
\end{equation} 
where
\begin{equation}\label{eq:R_def}
\textbf{R} =\begin{pmatrix}
\cos{\omega z} & -\sin{\omega z} \\
\sin{\omega z}  & \cos{\omega z} 
\end{pmatrix} 
\end{equation}
with $\omega$ being a real number. We will denote the two columns of the matrix $\mathbf{R}$ by $\mathbf{p}_1(z)$ and $\mathbf{p}_2(z)$. From \eqref{eq:A} it is clear that the eigenvalues of $\textbf{A}$ are $\lambda_1$ and $\lambda_2$ with the corresponding eigenvectors  being $\mathbf{p}_1(z)$ and $\mathbf{p}_2(z)$.

\subsection{Local stability of the invariant manifold}\label{sec:stability}
The Jacobian corresponding to the linearization of  \eqref{eq:diff_eq} at any point $(x,y,z)$ is
\begin{equation}\label{eq:J_def}
\textbf{J} = \left(\begin{array}{cc|c}
A_{11} &  A_{12} & x\frac{\partial A_{11}}{\partial z} + y\frac{\partial A_{12}}{\partial z}\\[0.6em]
A_{21} &  A_{22}& x\frac{\partial A_{21}}{\partial z} + y\frac{\partial A_{22}}{\partial z}\\[0.6em]
\hline
0 & 0 & \frac{\partial g}{\partial z}
\end{array} \right) 
\end{equation}
where $A_{ij}$ denotes the element of the matrix $\mathbf{A}$ on the $i$th row and $j$th column. The Jacobian $\mathbf{J}$ has a block upper triangular form
\begin{equation}\label{eq:J_form}
\textbf{J} = \left(\begin{array}{cc}
\mathbf{A} & \mathbf{B}_{2\times 1} \\[0.5em]
\mathbf{0}_{1 \times 2} & \frac{\partial g}{\partial z}
\end{array} \right) .
\end{equation}

The eigenvectors and eigenvalues of $\mathbf{J}$ can be computed easily due to the block  upper triangular form of  $\mathbf{J}$. A formal substitution shows that
\begin{equation}
\mathbf{w}_1(z) = \begin{pmatrix}
\mathbf{p}_1(z)\\
0
\end{pmatrix} \;\;\;\;\; 
\text{and} \;\;\;\;\;
\mathbf{w}_2(z) = \begin{pmatrix}
\mathbf{p}_2(z)\\
0
\end{pmatrix} \;\;\;\;\;
\end{equation}
are two of the eigenvectors of $\mathbf{J}$ with the corresponding eigenvalues being $\lambda_1$ and $\lambda_2$. 
\[
\mathbf{J}\begin{pmatrix}
\mathbf{p}_1\\
0
\end{pmatrix} = \left(\begin{array}{cc}
\mathbf{A} & \mathbf{B}_{2\times 1} \\[0.5em]
\mathbf{0}_{1 \times 2} & \frac{\partial g}{\partial z}
\end{array} \right)\begin{pmatrix}
\mathbf{p}_1\\
0
\end{pmatrix} = \begin{pmatrix}
\mathbf{A}\mathbf{p}_1\\
0
\end{pmatrix} = \lambda_1\begin{pmatrix}
\mathbf{p}_1\\
0
\end{pmatrix}.
\]
An analogous calculation can be performed for the second eigenvector $\mathbf{w}_2$. 

The sum of the eigenvalues of $\mathbf{J}$ is 
\[
tr(\mathbf{J}) = tr(\mathbf{A}) + \frac{\partial g}{\partial z} = \lambda_1+\lambda_2 +\lambda_3\]
Where, $tr(\mathbf{A})=\lambda_1+\lambda_2$. Hence, the third eigenvalue of $\mathbf{J}, \lambda_3$, is indeed $\frac{\partial g}{\partial z}$. We will  choose $g(z)$ such that 
\begin{equation}\label{eq:dg_dz}
\frac{\partial g}{\partial z} \neq 0 \;\;\;\; \text{for all } z.
\end{equation}

\begin{figure}[!h]
	\begin{minipage}{0.48\hsize}
		\begin{center}
			\includegraphics[width =\hsize]{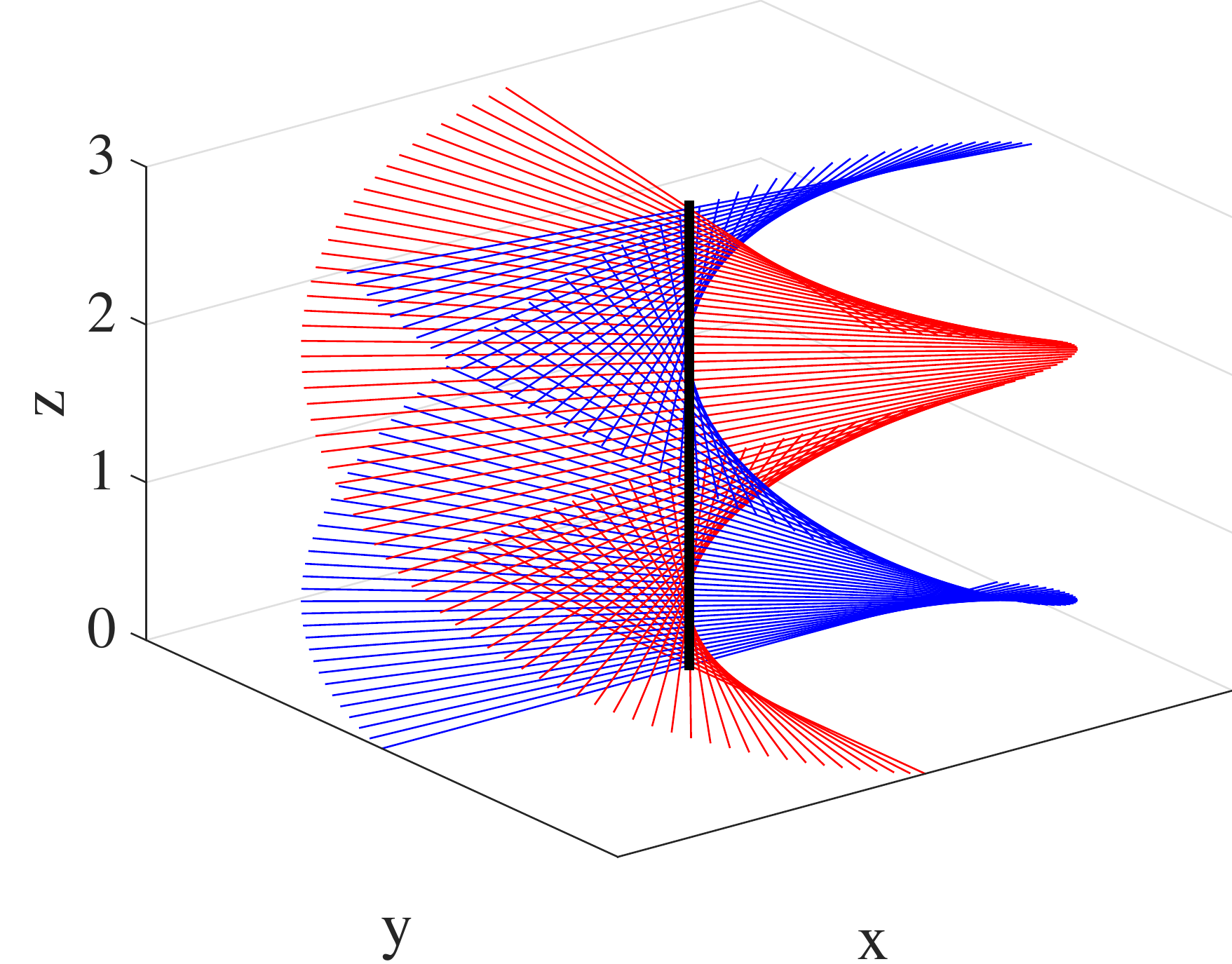}\\
		\end{center}
	\end{minipage}
	\begin{minipage}{0.48\hsize}
		\begin{center}
			\includegraphics[width =\hsize]{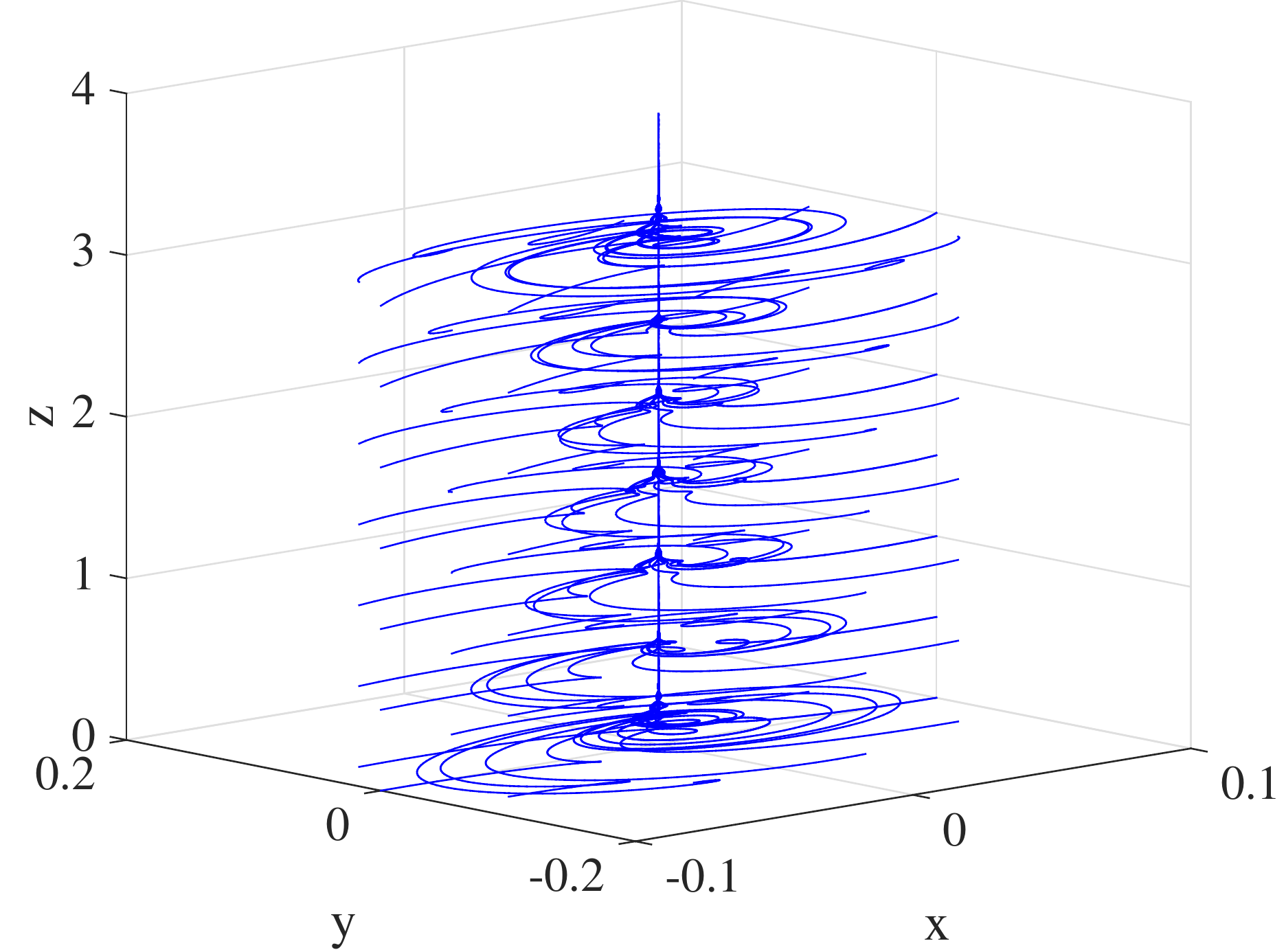}\\
		\end{center}
	\end{minipage}

	\caption{ (a)  Eigenvectors $\mathbf{w}_1$ and $\mathbf{w}_2$  spanning the stable and unstable subspaces  of $N_{\xi}M$ for several values of $(x=0,y=0,z)$. The (stable) vectors $\mathbf{w}_1$ are in blue and the (unstable) vectors $\mathbf{w}_2$ are in red. (b) Trajectories with various initial conditions spiral around the $z-$ axis and converge to it. The simulations results are for $\omega = 30$ and $g(z) = 0.01(1+\sin^2{z})$.}
	\label{fig:es_eu}
\end{figure}

At any $ (0, 0, z) \in M$, the Jacobian is

\begin{equation}
\textbf{J}(x=0,y=0,z) =  \left(\begin{array}{cc}
\mathbf{A} & \mathbf{0}_{2\times 1} \\[0.5em]
\mathbf{0}_{1 \times 2} & \frac{\partial g}{\partial z}
\end{array} \right) .
\end{equation}
The third eigenvector of the Jacobian, $\textbf{J}(x=0,y=0,z)$ is 
\begin{equation}\label{eq:w3}
\mathbf{w}_3 = \begin{pmatrix}
0\\
0\\
1
\end{pmatrix}.
\end{equation}
The eigenvectors $\mathbf{w}_1$, $\mathbf{w}_2$ and $\mathbf{w}_3$ are mutually orthogonal. The vectors $\mathbf{w}_1$ and $\mathbf{w}_2$ span the normal space of $M$, with $\mathbf{w}_1$ spanning the stable subspace of $N_{\xi}M$ and $\mathbf{w}_2$ spanning the unstable space of $N_{\xi}M$. The effect of the transformation \eqref{eq:A} is such that these vectors undergo a continuous rotation along $M$ as shown in fig. \ref{fig:es_eu}.

With \eqref{eq:dg_dz}, the Jacobian matrix resulting from the linearization of the dynamical system given by \eqref{eq:diff_eq} has only real eigenvalues. Furthermore we will choose 
\begin{equation}\label{eq:zdot}
\dot{z} = k(1+\sin^2{z}),
\end{equation}
where $k=0.01$. Therefore $|\frac{\partial g}{\partial z}| \ll |\lambda_1|$ and $|\frac{\partial g}{\partial z}| \ll |\lambda_2|$. This means  the dynamics on $M$ itself are slower than the dynamics normal to $M$, making $M$ a normally hyperbolic invariant manifold (NHIM). We note however that the global attractivity of $M$ can occur even if the dynamics on $M$ are not slow. From the definition of $A(z)$, $\lambda_2>0$ at all points $\xi = (0,0,z) \in M$, hence the NILE, $\sigma(\xi) = \lambda_2 >0$ everywhere on the invariant manifold. This implies that $M$ is locally unstable at all points $\xi\in M$. Any neighborhood of $\xi \in M$ has a subset that will be repelled away from $M$.

The relevance of the NILE as an indicator of local instability can be seen through a direct calculation to show that any neighborhood  of $z\in M$ contains a subset that is temporarily repelled from the invariant set.  A perturbation $\mathbf{u} =  \begin{pmatrix}x\\y\end{pmatrix}$ is decreasing in norm if the condition $\frac{d |\mathbf{u}|^2}{dt} < 0$ is met. Noting that $|\mathbf{u}|^2 = \mathbf{u}^T\mathbf{u}$ and $\dot{\mathbf{u}} = \mathbf{R}\mathbf{A}\mathbf{R}^T\mathbf{u}$, the rate of growth of a perturbation can be calculated as follows
\begin{equation}\label{eq:rate}
 \frac{d \mathbf{u}^T\mathbf{u}}{dt} = (\mathbf{R}\mathbf{A}\mathbf{R}^T \mathbf{u})^T\mathbf{u} + \mathbf{u}^T\mathbf{R}\mathbf{A}\mathbf{R}^T\mathbf{u} = 2\mathbf{u}^T\mathbf{B}\mathbf{u}
\end{equation}
where $\mathbf{B} = \mathbf{R}\mathbf{A}\mathbf{R}^T$. If every perturbation in a neighborhood of $z \in M$ decays, then $S(z)<0$, which requires that $\mathbf{B}(z)$ be negative definite. Since $\mathbf{B}$ is obtained from $\mathbf{A}$ through rotations, the trace and determinant of $\mathbf{B}$ are the same as that of $\mathbf{A}$. This can of course be verified through a direct calculation.
\[
\mathbf{B}= \begin{pmatrix}
\lambda_1\cos^2{\omega z}+\lambda_2\sin^2{\omega z} & \cos{\omega z}\sin{\omega z}(\lambda_1-\lambda_2) \\
\cos{\omega z}\sin{\omega z}(\lambda_1-\lambda_2)  & \lambda_2\cos^2{\omega z}+\lambda_1\sin^2{\omega z}
\end{pmatrix}.
\] 
Since we are considering the case where the eigenvalues of $\mathbf{A}$ are real and of opposite sign, with the negative eigenvalue having the larger magnitude,  we obtain  $tr(\mathbf{B }) = \lambda_1+\lambda_2 <0$ and $det(\mathbf{B}) = \lambda_1\lambda_2<0$.   Therefore $\mathbf{B}(z)$ is not negative definite for any value of $z$. Therefore any neighborhood of $z \in M$ has a subset that can grow in norm, at least temporarily and the $z-$ axis is locally unstable everywhere.

\begin{figure}[!h]
	\begin{minipage}{0.3\hsize}
		\begin{center}
			\includegraphics[width =\hsize]{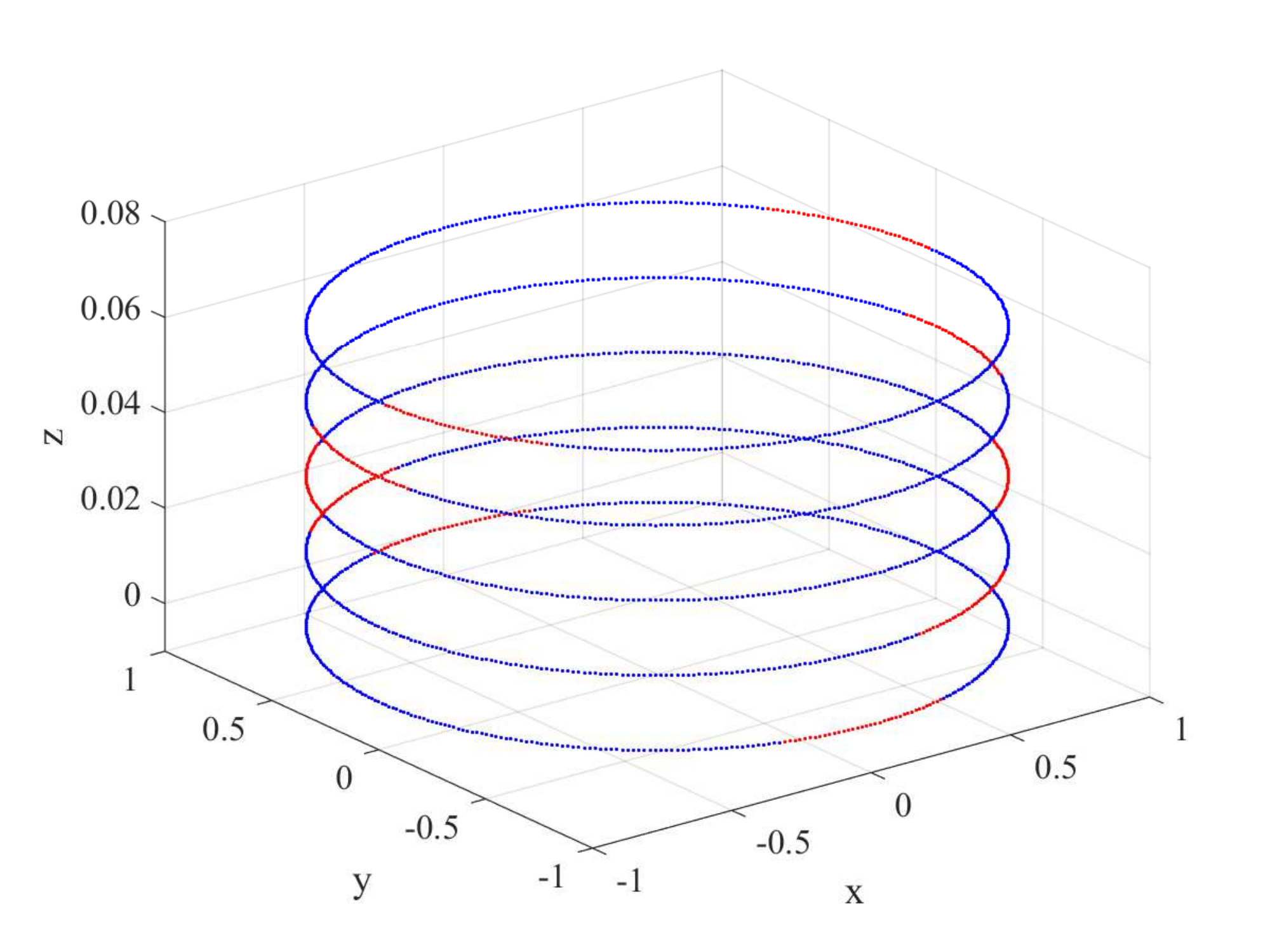}\\
(a) 
		\end{center}
	\end{minipage}
	\begin{minipage}{0.3\hsize}
		\begin{center}
			\includegraphics[width =\hsize]{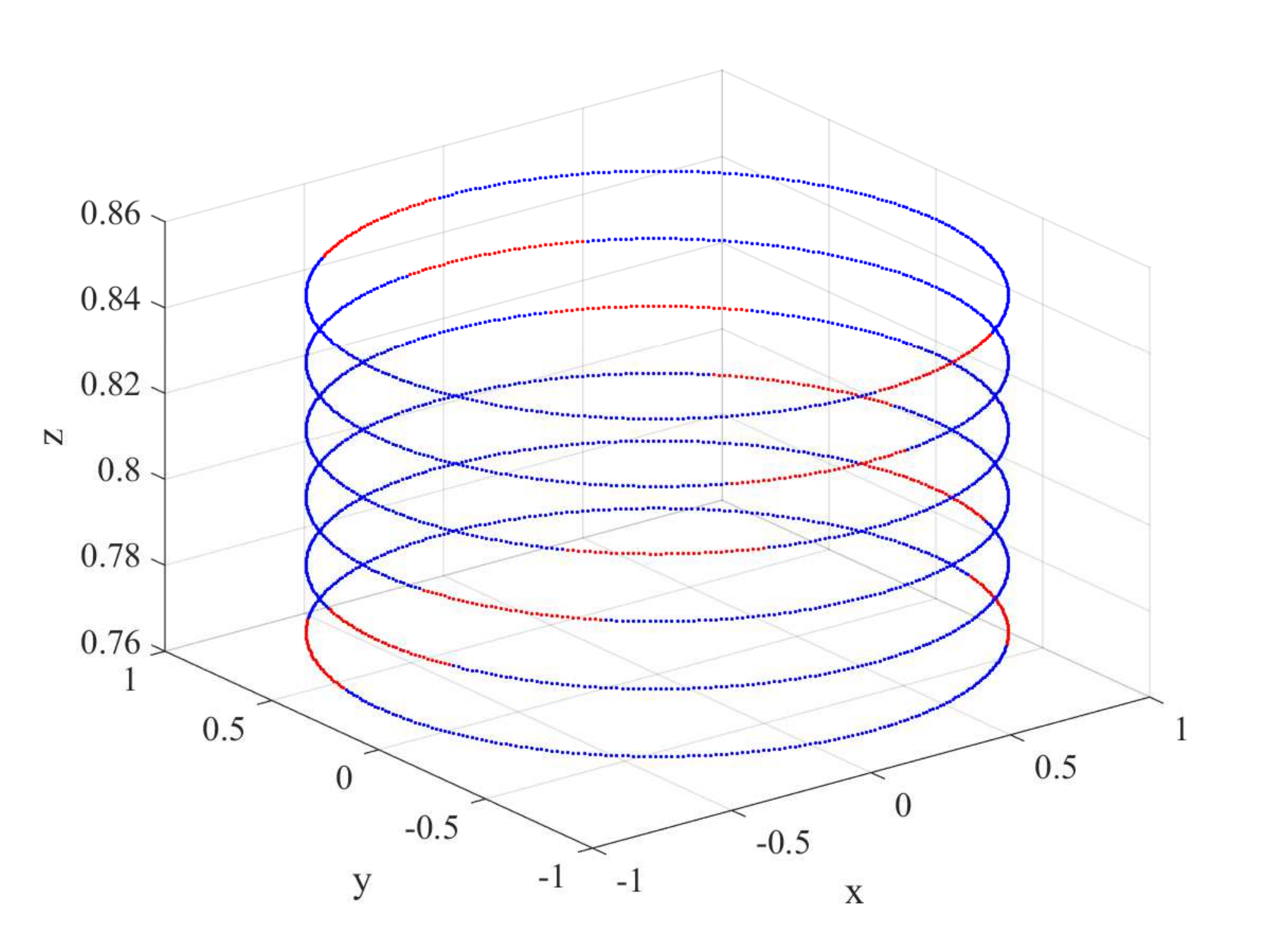}\\
(b)
		\end{center}
	\end{minipage}
	\begin{minipage}{0.3\hsize}
		\begin{center}
			\includegraphics[width =\hsize]{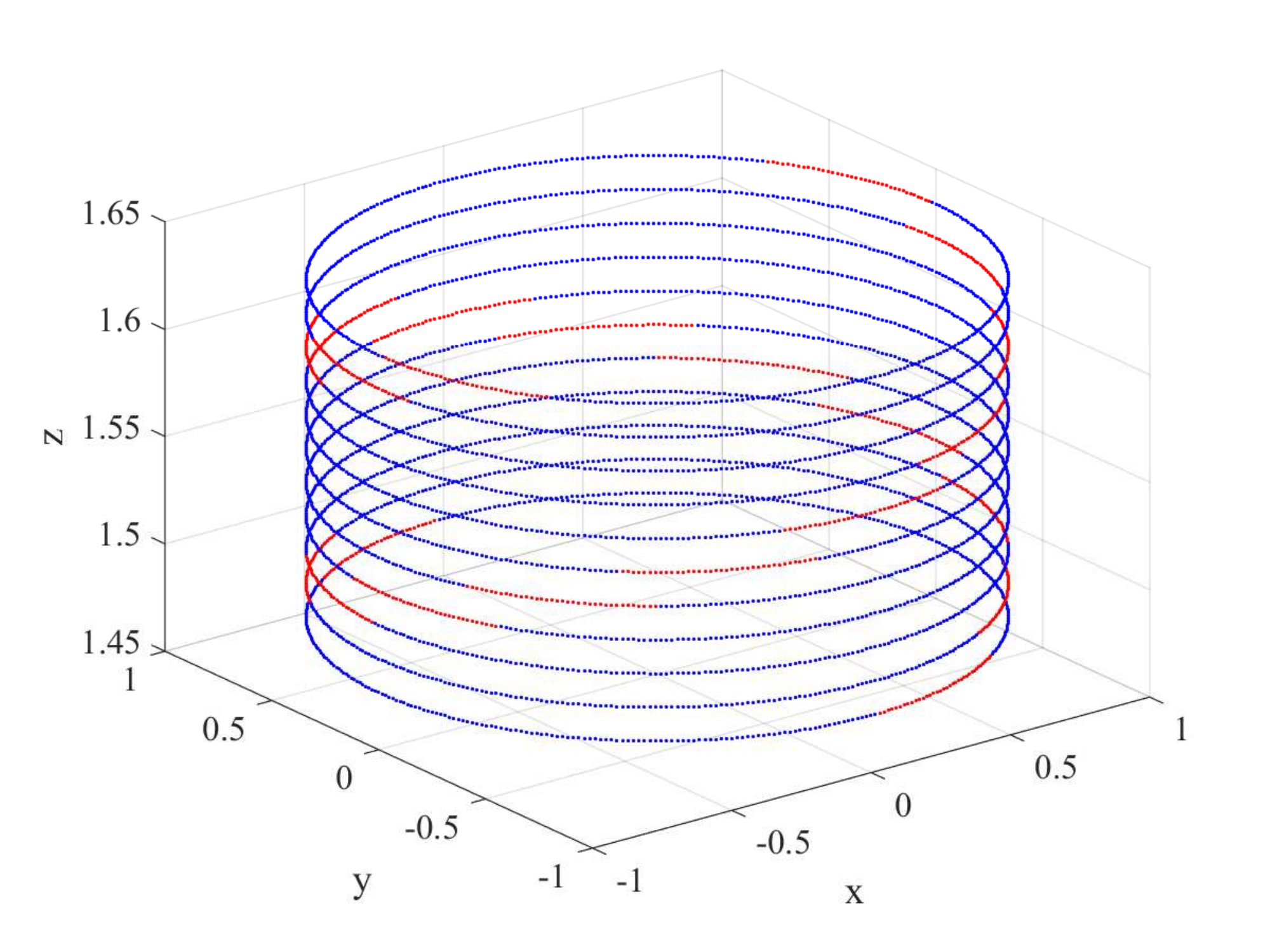}\\
(c) 
		\end{center}
	\end{minipage}

	\caption{ Unit circles around the $z-$ axis are selected for different ranges of $z$. (a) $z \in [0,0.02\pi]$, (b) $z \in [0.2450\pi,0.27\pi]$ and (c) $z \in [0.47\pi, 0.52\pi]$. Perturbations on the blue subsets decay instantaneously ($S<0$) while those on the red subsets grow instantaneously ($S>0$). Such locally unstable subsets of a neighborhood are present for any value of $z$.} \label{fig:subsets}
\end{figure}

Suppose one chooses perturbations on a circle centered at some $z$, $\mathbf{u} = r(\cos{\theta}, \sin{\theta})$, and denoting $S(z,r,\theta) = 2\mathbf{u}^T\mathbf{B}\mathbf{u}$,
\begin{align}\label{eq:S}
S(z, r, \theta) =&2 r^2(B_{11}\cos^2{\theta} + B_{22} \sin^2{\theta} \nonumber \\ & + (B_{12}+B_{21})\cos{\theta}\sin{\theta}) 
\end{align}
where $B_{ij}$ represent the entries in the matrix $\mathbf{B}$. The sign of $S$ 
can be found numerically for different values of $\theta$ for a given $z$. The locally stable subsets of a circular neighborhood of $z$ are $S^-(z) = \{\theta: S(z,r,\theta) <0\}$, while the locally unstable subsets of the circular neighborhood are $S^+(z) = \{\theta: S(z,r,\theta) >0\}$. Figure \ref{fig:subsets} shows the stable (blue) and unstable (red) subsets of a unit circle for different ranges of $z$ along the invariant manifold.  

 At the origin
\begin{equation}\label{eq:S0}
S(0, r, \theta) = 2r^2(\lambda_1\cos^2{\theta} + \lambda_2\sin^2{\theta})
\end{equation}
and $S^-(0)$ is the set that satisfies,
\begin{equation}\label{eq:condition}
\tan^2{\theta} < -\frac{\lambda_1}{\lambda_2}.
\end{equation}
The stable and unstable subsets on the unit circle at any value of $z$ are obtained through a rigid rotation of these sets when $z=0$. The relative measures of the unstable and stable subsets on a unit circle remain constant at any point on the $z$-axis. 

Since the rate of growth of $|\mathbf{u}|^2$ instead of $\mathbf{u}$ was computed in \eqref{eq:rate}, the rate of growth of the perturbation itself is given by $\frac{S}{2r}$. Furthermore normalizing this by the norm of the initial perturbation, gives  $\frac{S}{2r^2}$. The relationship of $S$ to the NILE can be understood by computing the maximum value of $\frac{S}{2r^2}$. By inspecting \eqref{eq:S0}, $max(\frac{S}{2r^2})= \lambda_2$, since $\lambda_1<0<\lambda_2$.  Therefore $max(\frac{S}{2r^2}) = \sigma$, the NILE. 

Here we comment on an alternative way to characterize the local stability   of a limit cycle, such as in \cite{ali_menzinger_chaos99}. A transformation of coordinates that rotates with the eigenvectors of $\mathbf{A}$ along $M$ can be performed, to obtain $\mathbf{v} = \mathbf{R}(\omega z) \mathbf{u}$. Differentiating this expression with time one obtains
\begin{equation}\label{eq:rot}
\frac{d\mathbf{u}}{dt} = (\mathbf{A} - \mathbf{\Omega}\dot{z}) \mathbf{u}.
\end{equation}
Here we use the fact 
\[
\frac{d \mathbf{R}(\omega z)}{dt} = \mathbf{\Omega} \mathbf{R} \dot{z}
\]
where $\mathbf{\Omega}$ is the skew-symmetric matrix
\[
\mathbf{\Omega} = \begin{pmatrix}
0 & -\omega\\
\omega & 0
\end{pmatrix}.
\]
In the event that $\dot{z}$ is constant, \eqref{eq:rot} is independent of the flow on $z$-axis. The stability of the fixed point of \eqref{eq:rot}, $\mathbf{u}=0$ can be decided by the eigenvalues of $\mathbf{C}= (\mathbf{A} - \mathbf{\Omega}\dot{z})$. If both the eigenvalues of $\mathbf{C}$ then all perturbations will decay. The eigenvalues of $\mathbf{C}$ are negative for a high enough $\omega$. While this transformation to a rotating frame of reference is convenient, the local stability of $M$ cannot be inferred from the eigenvalues of $\mathbf{C}$ if $\dot{z}$ is not constant. Even when $\dot{z}$ is constant, it can be seen that some perturbations can temporarily grow,
\begin{equation}\label{eq:rot_growth}
\frac{d \mathbf{u}^T\mathbf{u}}{dt} = (\mathbf{C} \mathbf{u})^T\mathbf{u} + \mathbf{u}^T\mathbf{C}\mathbf{u} = \mathbf{u}^T(\mathbf{C}^T+\mathbf{C})\mathbf{u} = 2 \mathbf{u}^T\mathbf{A}\mathbf{u}.
\end{equation}
The right hand side of \eqref{eq:rot_growth} is the same as that of \eqref{eq:S0} when $\omega z=0$. From the previous analysis there is a finite subset of perturbations that grow temporarily. This local instability is in fact independent of $\omega$, but depends only on $\lambda_1$ and $\lambda_2$. Drawing conclusions of local stability based on the eigenvalues of $\mathbf{C}$ underestimates the regions of instability and the parameter space of instability. Such under estimates of domains of local instability based on the eigenvalues of a Jacobian matrix have been made in other contexts. For instance \cite{babiano} provided a lower estimate for the regions of instabilities in a fluid domain where inertial particles can deviate from streamlines and the correct calculations using the NILE in \cite{haller_pof_2008} showed the regions of local instabilities to be much larger.

\subsection{Global stability of the invariant manifold}
While small perturbations normal to $M$ could be repelled away for intermediate periods of time, the long time behavior of the trajectories is more complex. Numerical simulations of  trajectories, see fig. \ref{fig:es_eu}(b), with several initial conditions $(x\neq0,y\neq0,z)$ demonstrate that  all trajectories of \eqref{eq:diff_eq} converge to $M$, (`z-axis'). Numerical simulations of trajectories starting very far from the $z-$axis ($d(0) = 10^6$) also show that they converge to the invariant manifold. The values of the parameters of the system that are chosen for these simulations are $\lambda_1 = -1.1$, $\lambda_2 = 0.1$ and $\omega = 25$. The numerical integration is performed in MATLAB using the solver \textit{ode113} which is a multistep variable order Adams Bashworth Moulton solver with an error tolerance of $10^{-14}$. The numerical integration was performed for large time periods ranging from $t=10^5$ to $t=10^7$ to confirm the eventual non divergence of trajectories.

\begin{figure}[!h]
\begin{minipage}{0.48\hsize}
		\begin{center}
			\includegraphics[width =\hsize]{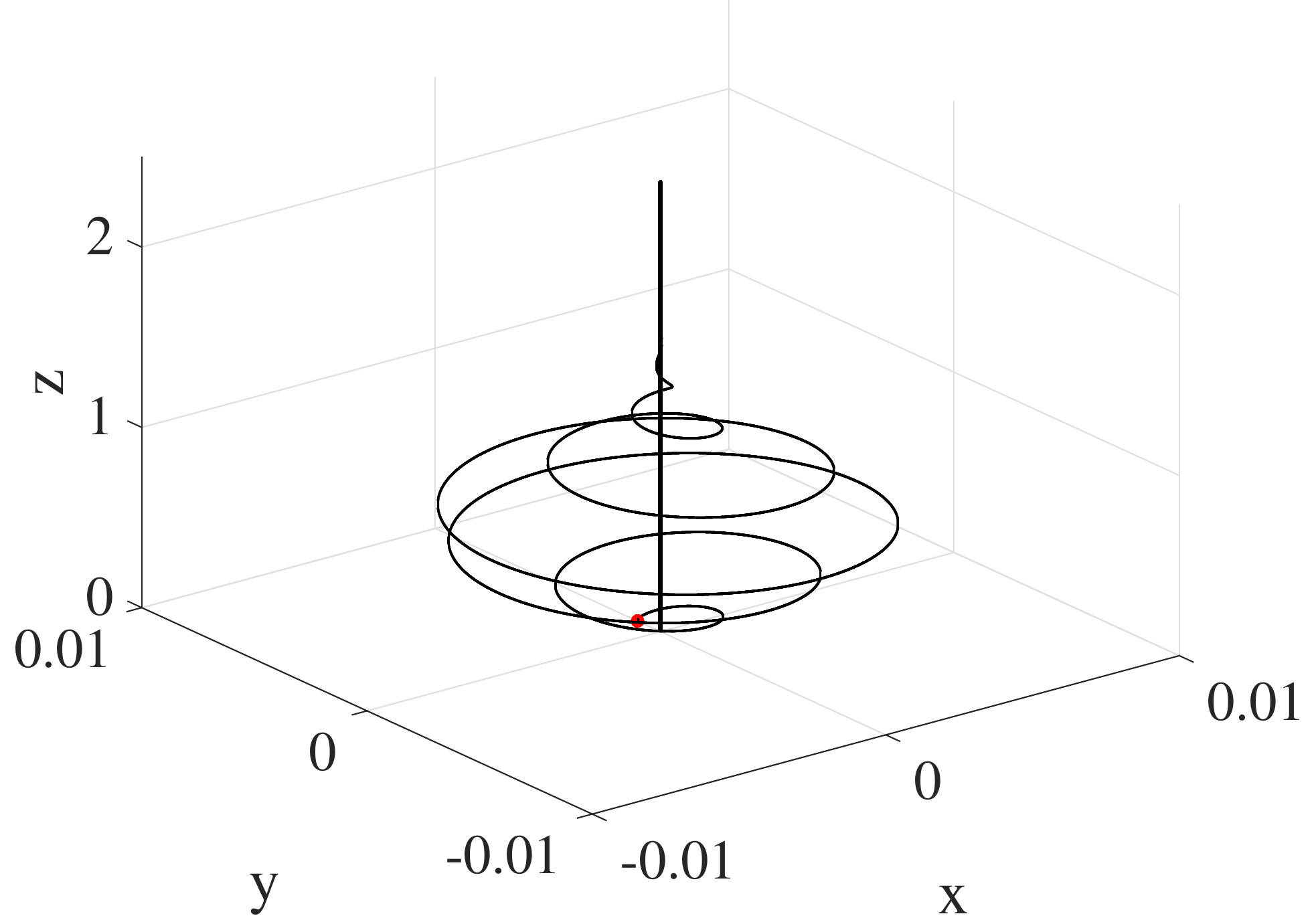}\\
(a)
		\end{center}
	\end{minipage}
	\begin{minipage}{0.48\hsize}
		\begin{center}
			\includegraphics[width =\hsize]{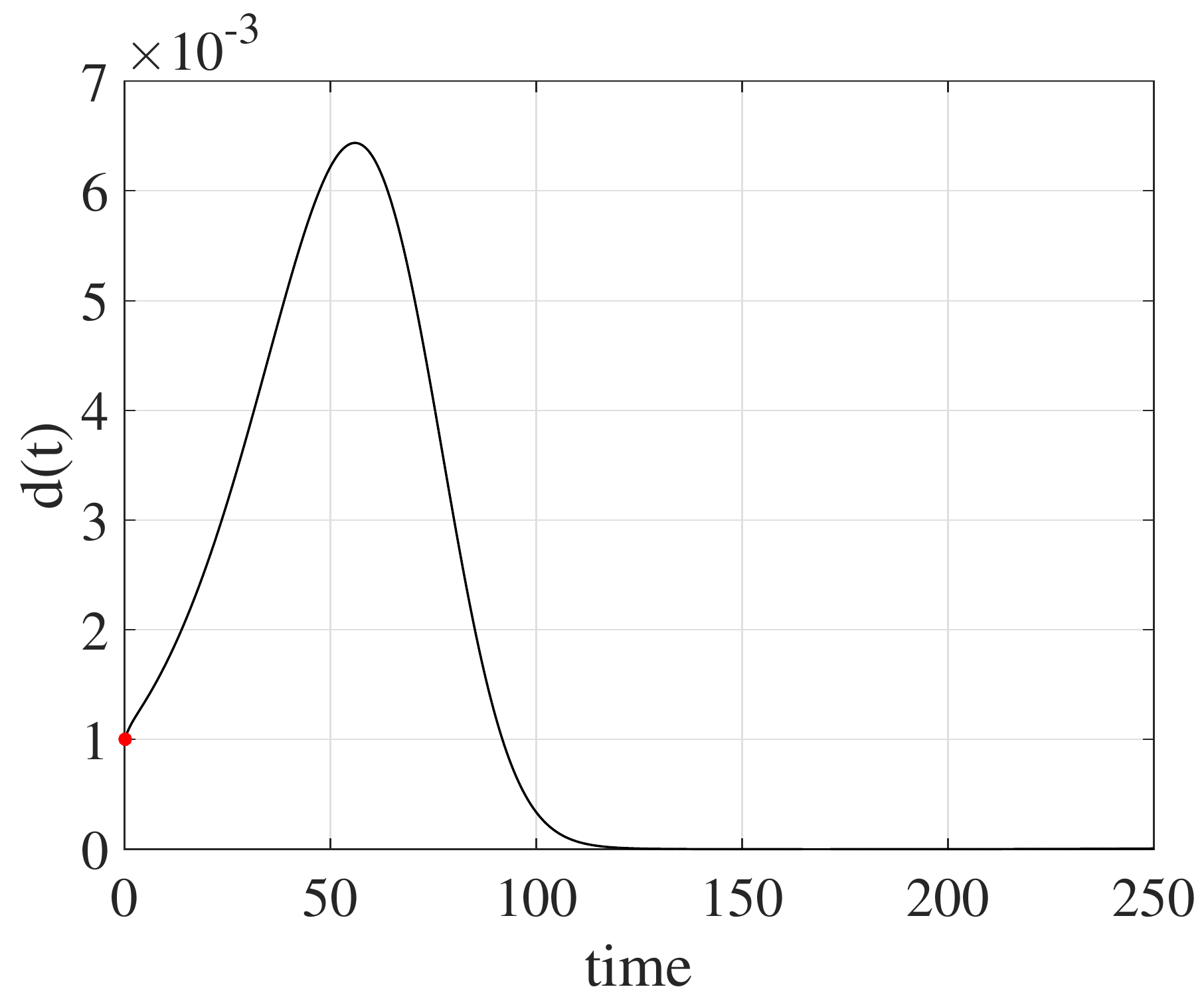}\\
(b)
		\end{center}
	\end{minipage}
\caption{ Perturbation along the unstable direction normal to $M$ at $z=0$. The initial conditions $(x(0) = 0, y(0) =0.001)$. (a) Trajectory in the $x-y$ plane, with the (red) circle showing the initial condition. (b) The distance $d(t)$ between the trajectory and the invariant manifold, $M$.  $\omega=25$.}\label{fig:unstable_perturb}
\end{figure}

 These numerics are  discussed here first for two special cases; a perturbation along locally unstable directions at the origin, with $(x(0) = 0.0, y(0) = 0.001, z=0)$ and a perturbation along the locally stable direction at the origin with $(x(0) = 0.001, y(0) = 0, z=0)$. Figure \ref{fig:unstable_perturb}(a) shows the trajectory and \ref{fig:unstable_perturb}(b) shows the distance $d(t)$ of the trajectory from the $z-$ axis. Since the perturbation is along the locally unstable direction at the origin, the distance of the trajectory from the $z-$axis initially increases more than six fold. However as the trajectory simultaneously spirals around the $z-$axis, its distance from the $z$- axis decreases and the trajectory converges to the $z-$ axis.
 
\begin{figure}[!h]

\begin{minipage}{0.48\hsize}
		\begin{center}
			\includegraphics[width =\hsize]{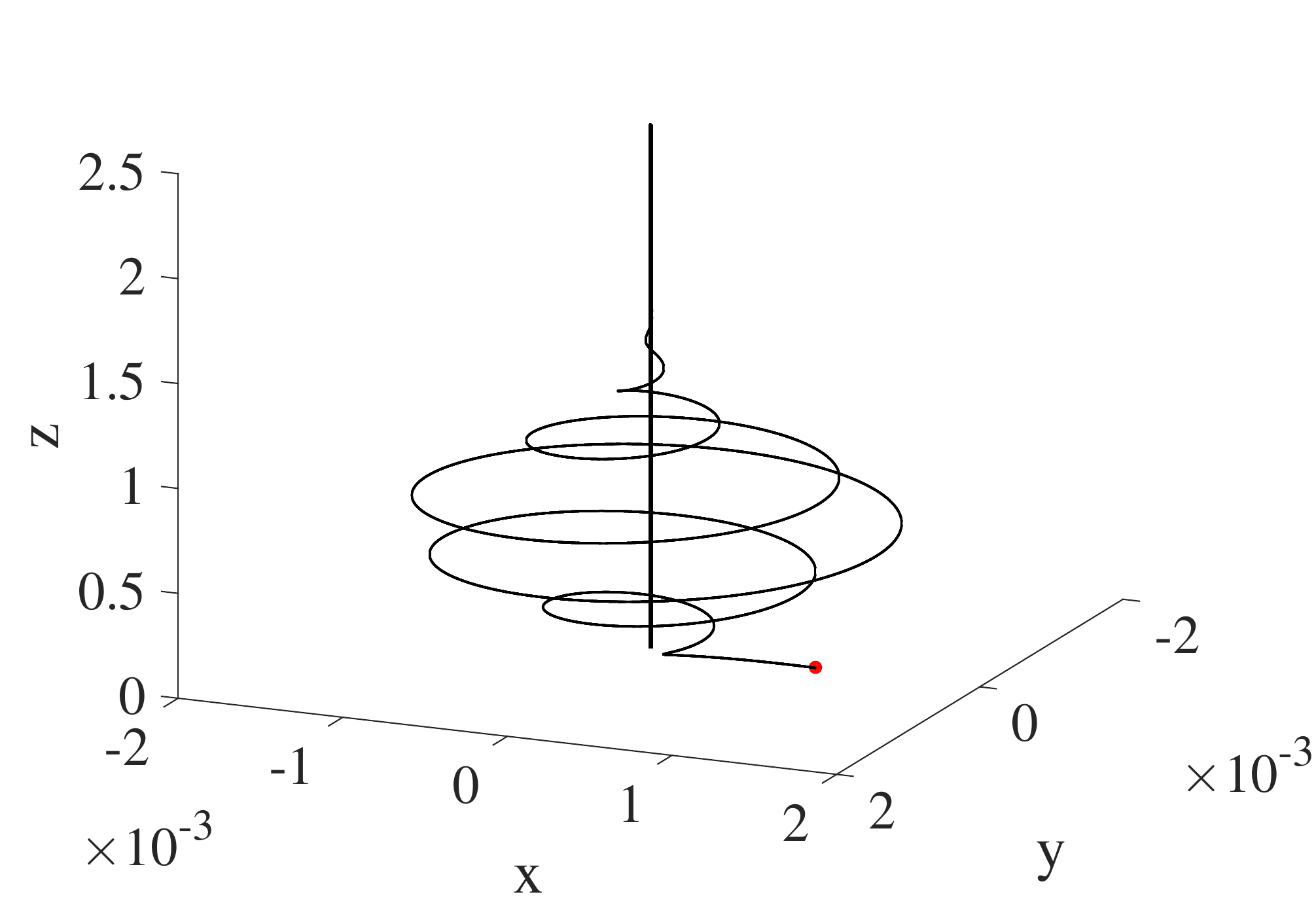}\\
(a)
		\end{center}
	\end{minipage}
	\begin{minipage}{0.48\hsize}
		\begin{center}
			\includegraphics[width =\hsize]{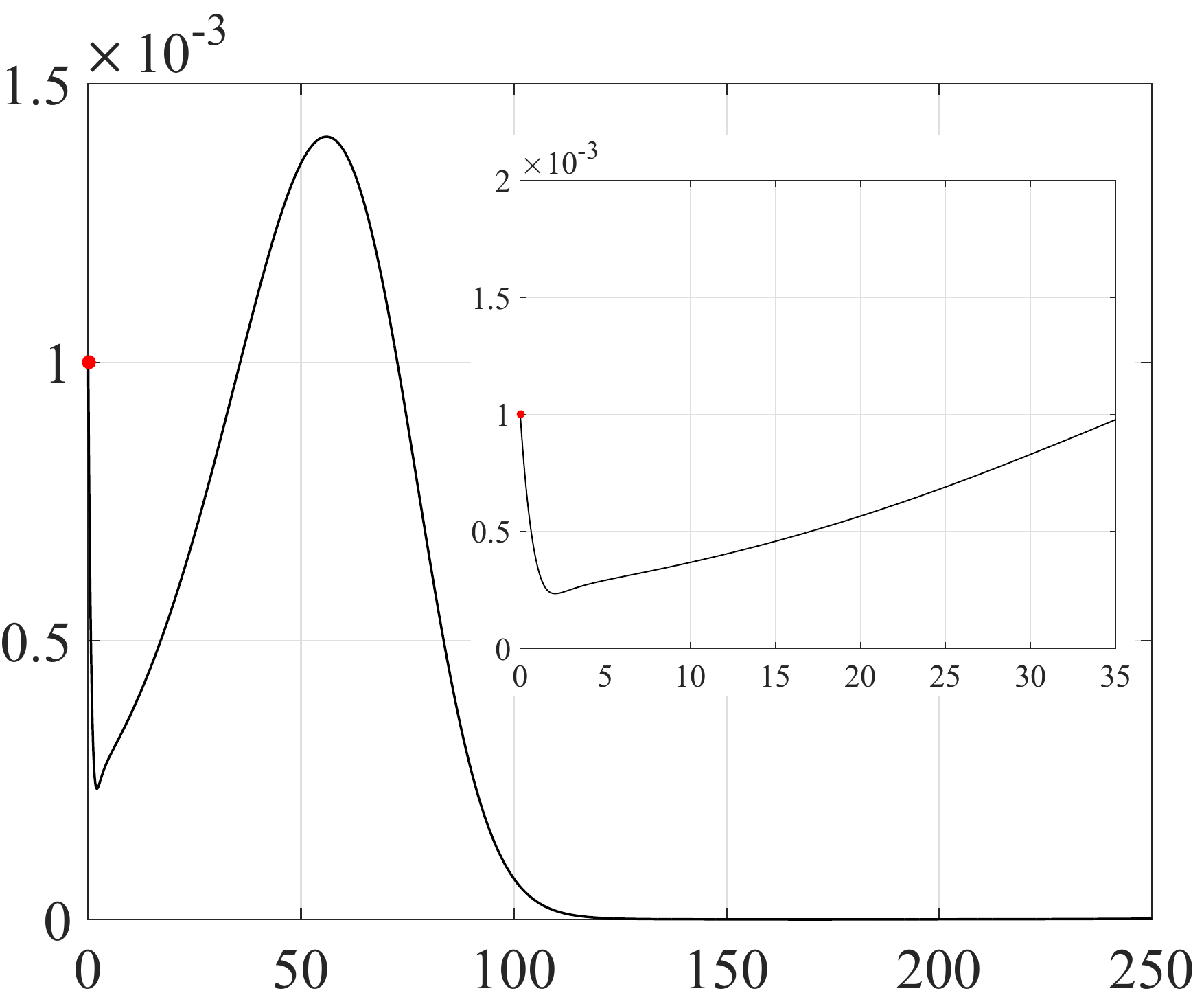}\\
(b)
		\end{center}
	\end{minipage}
\caption{ Perturbation along the stable direction normal to $M$ at $z=0$. The initial conditions $(x(0) = 0.001, y(0) =0)$. (a) Trajectory in the $x-y$ plane, with the (red) circle showing the initial condition. (b) The distance $d(t)$ between the trajectory and the invariant manifold, $M$ first decreases, then increases and then eventually decreases again.  A magnified version of the distance graph for $t \leq 35$ is also shown. $\omega=25$.}\label{fig:stable_perturb}
\end{figure}

Figure \ref{fig:stable_perturb}(a) shows the  trajectory and \ref{fig:stable_perturb}(b) shows the distance $d(t)$ of the trajectory from the $z-$ axis. Since the perturbation is along the locally stable direction at the origin, the distance of the trajectory from the $z-$ axis initially decreases. However as the trajectory simultaneously spirals around the $z$-axis, it begins to be repelled. This repulsion increases the distance of the trajectory from the $z-$ axis before it eventually decreases again and converges to zero.

The surprising decay of normal perturbations can be understood through a simpler example; a planar dynamical system with a saddle type fixed point at the origin,
\begin{align} \label{eq:saddle}
\dot{x}_1 &= \mu_1x_1 \nonumber \\
\dot{x}_2 &= \mu_2x_2
\end{align}
where $\mu_1<0$ and $\mu_2>0$. If the initial conditions are $(x_1(0), x_2(0))$, the distance of the trajectory from the origin is $D(t) = \sqrt{x_1(0)^2e^{2\mu_1t} + x_2(0)^2e^{2\mu_2t}}$. Rescaling the initial conditions as $x_2(0) = \epsilon x_1(0)$, the distance is 
\begin{equation}
D(t) = x_1(0)e^{\mu_1t} \sqrt{1 + \epsilon^2e^{2(\mu_2-\mu_1)t}}.
\end{equation}
 When $\epsilon =0$, $D(t)$ decays to zero. For a small enough $\epsilon>0$ the distance of the trajectory to the origin first decays, in an interval $(0,T)$ before increasing monotonically for $t>T$. The critical value of $\epsilon$ for which the $D(t)$ is an increasing function at $t=0$ can be obtained by setting $\frac{d D^2(t)}{dt}|_{t=0} > 0$. This gives the critical value of $\epsilon$,
 \begin{equation}\label{eq:epsilon}
\epsilon_{cr}^2 = -\frac{\mu_1}{\mu_2}.
\end{equation}
If $\frac{y(t)}{x(t)} < \epsilon_{cr}$ then $D(t)$ is a decreasing function. When $\frac{y(t)}{x(t)} = \epsilon_{cr}$ $D(t)$ is a minimum and increases thereafter.  It should be noted that $\epsilon$ in \eqref{eq:epsilon} is the same as $\tan{\theta}$ in \eqref{eq:condition} and the two equations are equivalent in identifying the locally unstable subsets of a neighborhood of the invariant sets  \eqref{eq:diff_eq} and \eqref{eq:saddle}. 

\begin{figure}[!h]
		\begin{center}
			\includegraphics[width =\hsize]{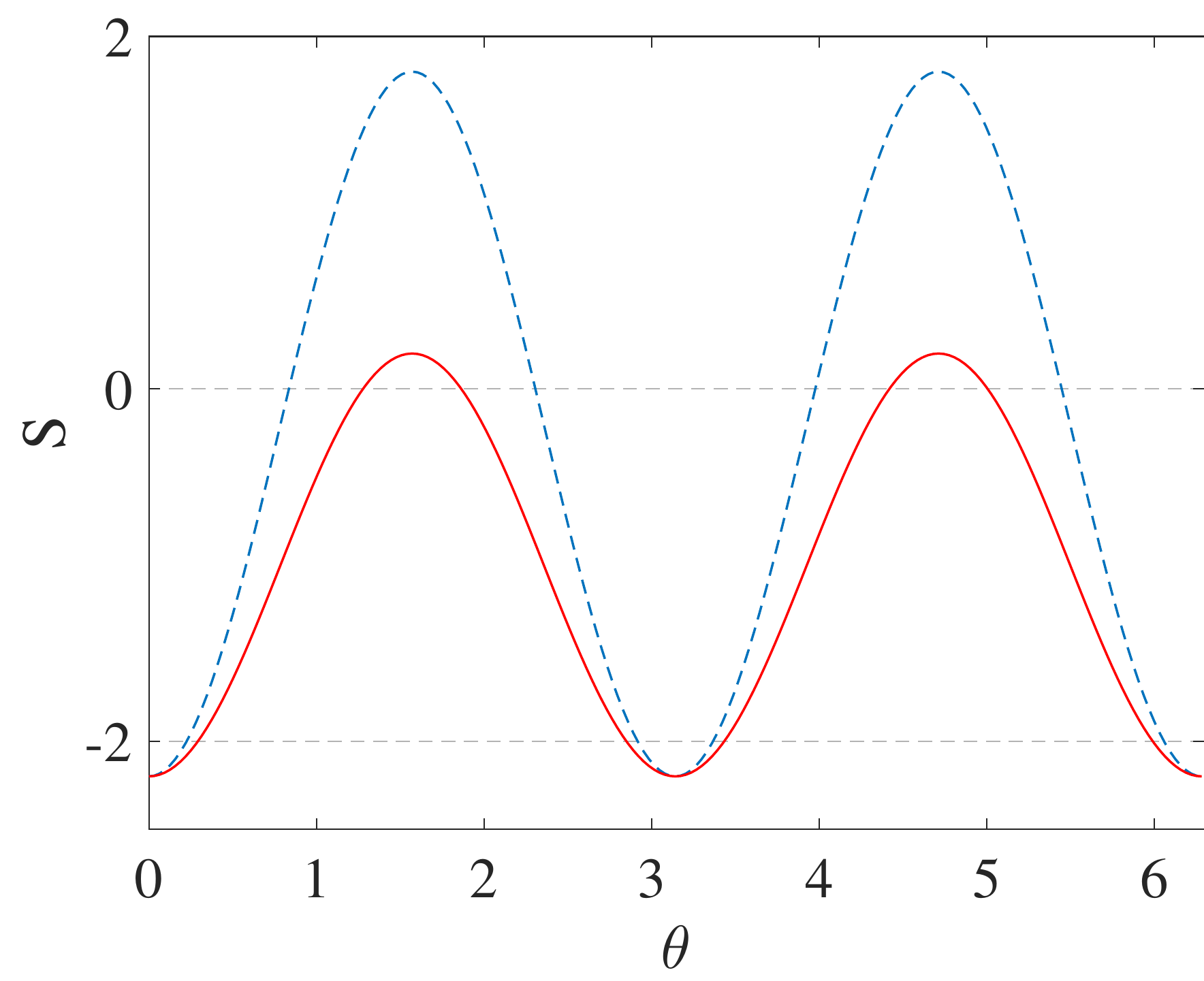}\\
		\end{center}
\caption{Graph of $S(0,r=1,\theta)$. (a) The solid red graph is for $\lambda_1=-1.1$, $\lambda_2 = 0.1$ and $\omega = 25$. (b) The blue dotted graph is for the case $\lambda_1 = -1.1$, $\lambda_2 = 0.9$ and $\omega = 98$.}\label{fig:S}
\end{figure}

We will denote the subset of $\mathbb{R}^2$ where $D(t)$ is decreasing, by $B_s$ and call it the decay set,
\begin{equation}
B_s = \{(x,y) | \frac{d}{dt}D(t) <0\}.
\end{equation} 
The sets $B_s \subset \mathbb{R}^2$ and $S^-$ are both trivially related; $S^-$ is merely the arc of a circle contained in the set $B_s$. The rate of growth of perturbations, $S$, \eqref{eq:S} is graphed in fig. \ref{fig:S} for two sets of parameters. The red dotted graph is for the case $\lambda_1=-1.1$, $\lambda_2 = 0.1$ and $\omega = 25$ and the blue dotted graph is for the case $\lambda_1 = -1.1$, $\lambda_2 = 0.9$ and $\omega = 98$. In both cases  $\lambda_1<0$ and $\lambda_1>\lambda_2$. This leads to the arc length of $S^-$ to be greater than $S^+$ and the maximum magnitude of decay to be larger than the maximum magnitude of repulsion.

The role of the saddle is played by the $z$-axis in the dynamical system \eqref{eq:diff_eq}. The difference is that the stable and unstable subspaces undergo a continuous rotation along $z-$ axis. A trajectory $(x(t), y(t), z(t)) = \xi(t;t_0, \xi_0)$ moves closer to the invariant manifold when $(x(t), y(t)) \in B_s$. The decay set too undergoes a rotation along the $z-$ axis. When perturbed trajectories enter this decay set their distance to the $z-$axis decreases.

\begin{figure}[!h]

\begin{minipage}{0.48\hsize}
		\begin{center}
			\includegraphics[width =\hsize]{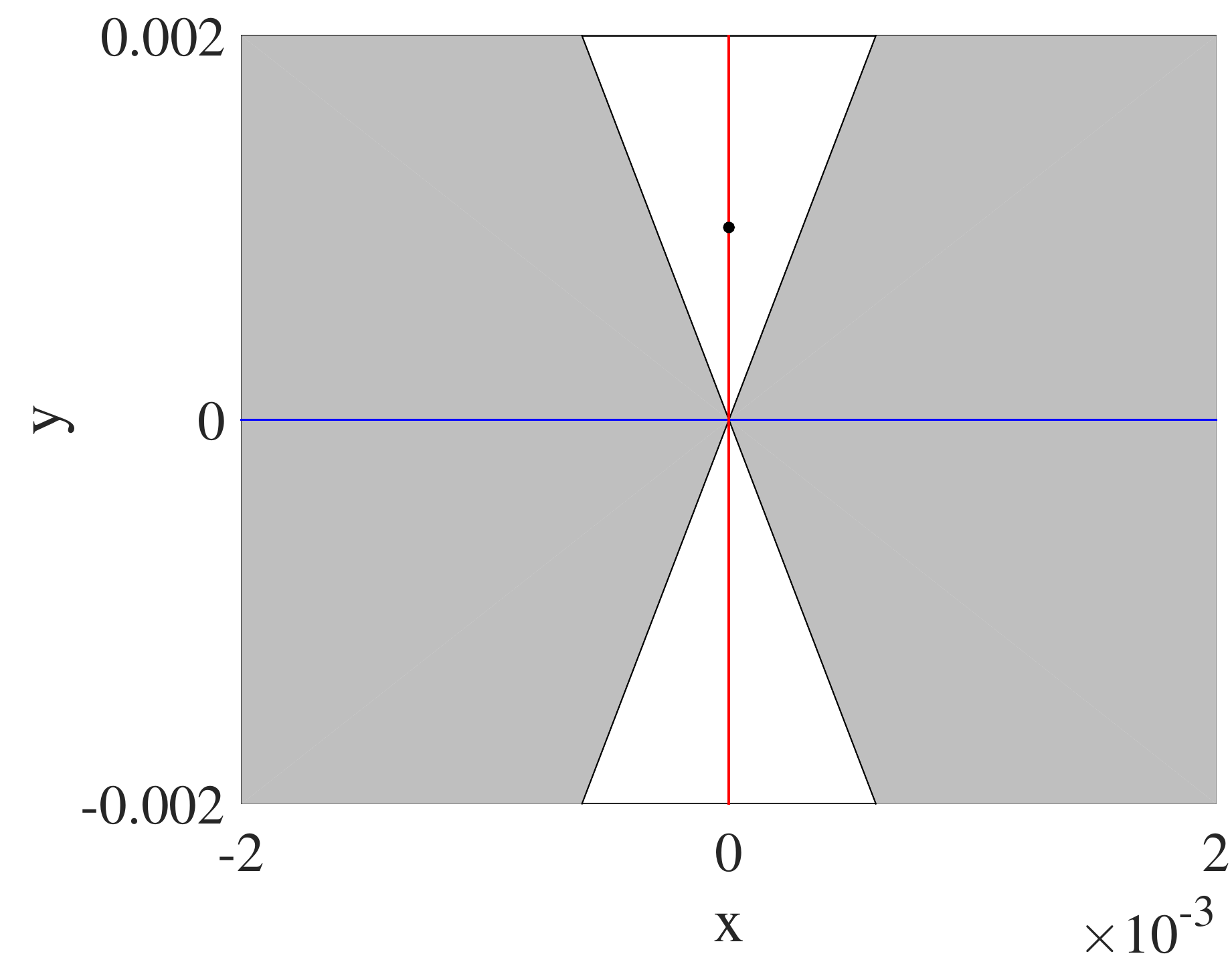}\\
(a) $t= 1.5$
		\end{center}
	\end{minipage}
	\begin{minipage}{0.48\hsize}
		\begin{center}
			\includegraphics[width =\hsize]{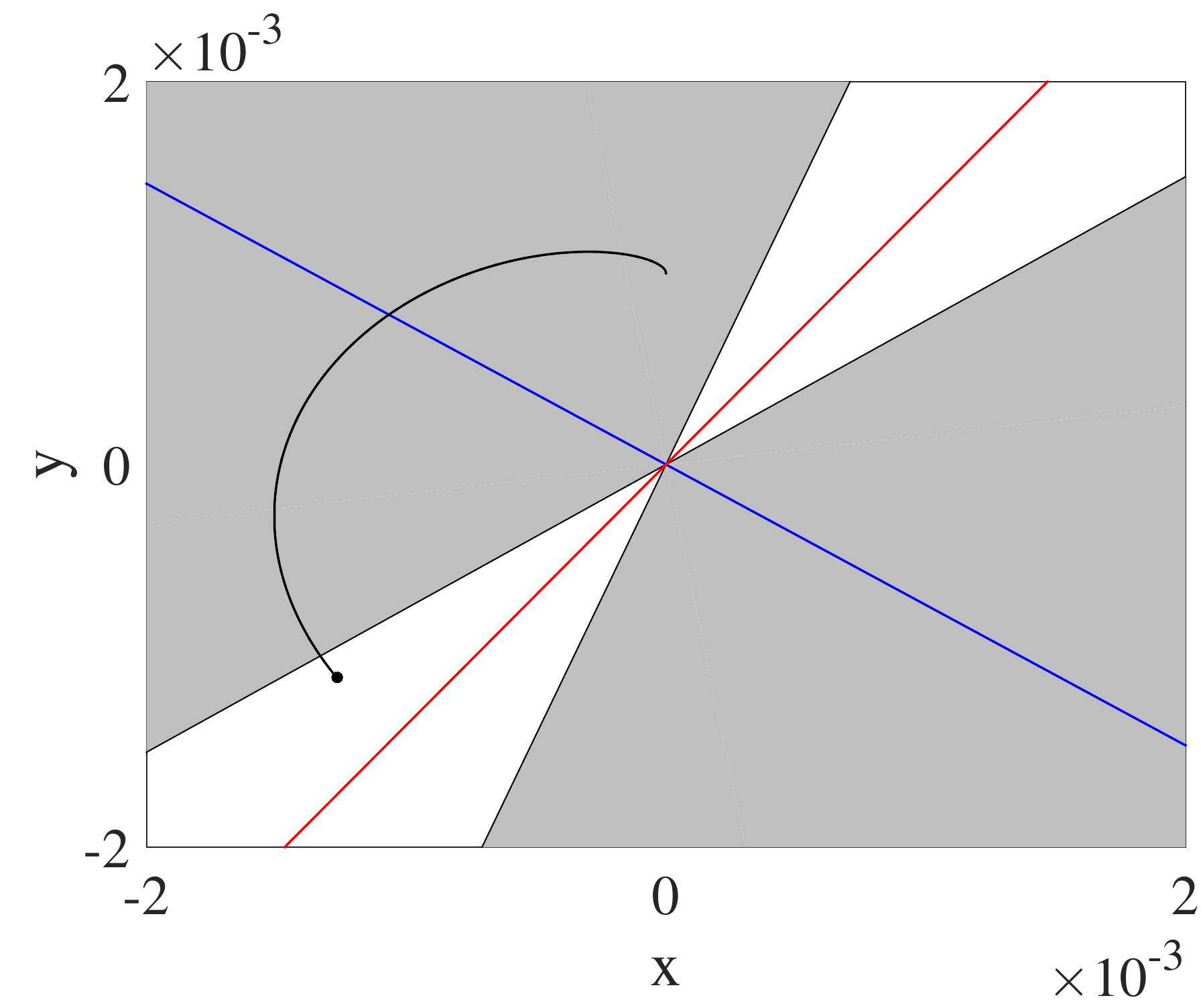}\\
(b) $t=10$
		\end{center}
	\end{minipage}

\begin{minipage}{0.48\hsize}
		\begin{center}
			\includegraphics[width =\hsize]{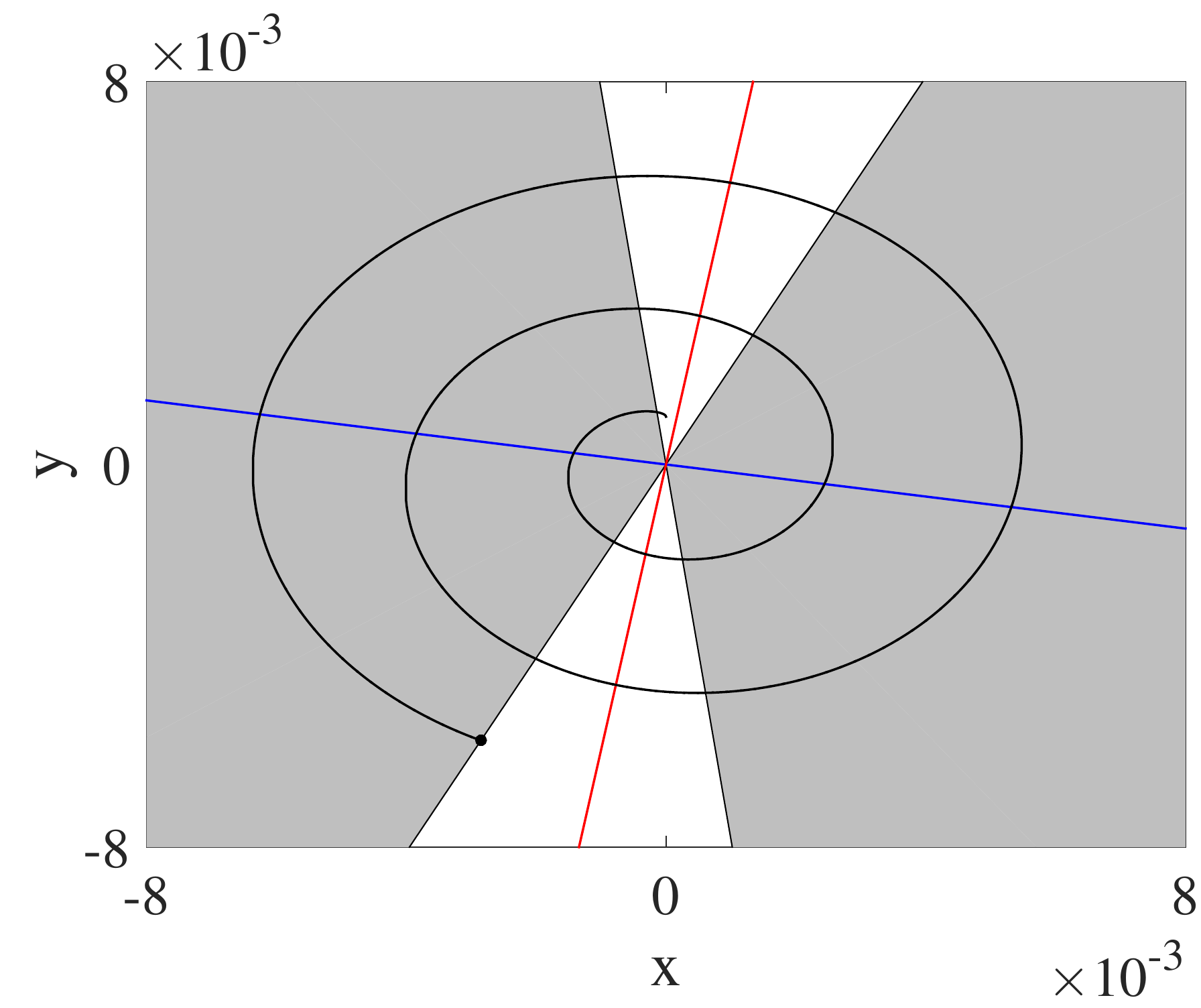}\\
(c) $t= 56$
		\end{center}
	\end{minipage}
	\begin{minipage}{0.48\hsize}
		\begin{center}
			\includegraphics[width =\hsize]{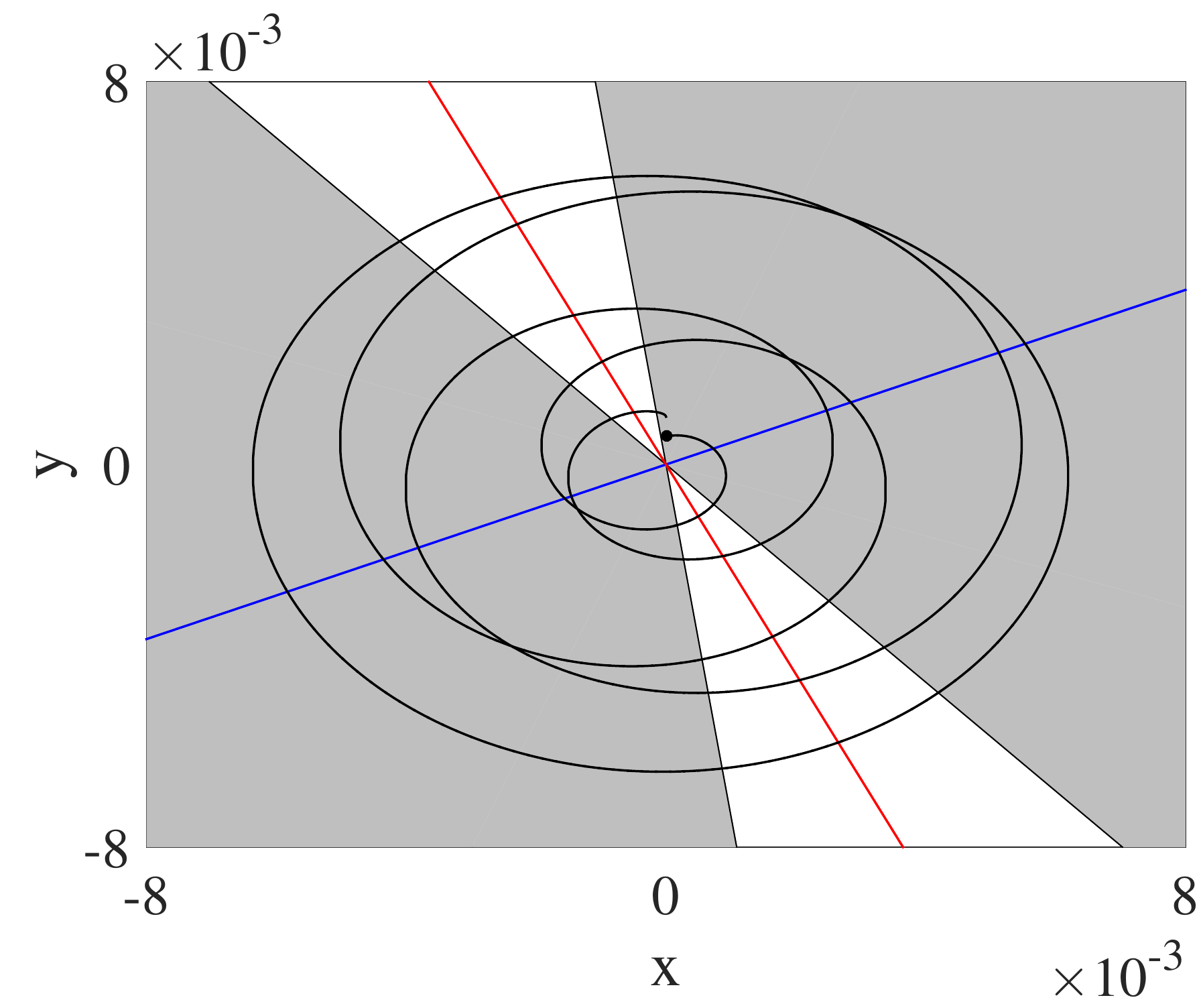}\\
(d) $t=96$
		\end{center}
	\end{minipage}
\caption{ The projection of the trajectory in the $x-y$ plane for $(x(0)=0, y(0) = 10^{-3})$. The decay set is shown in gray. In (a) and (b) the trajectory at $(x(t), y(t))$ (shown by a black circle) does not lie in the decay set while in (c) and (d) it does. }\label{fig:decay_set} 
\end{figure}

The decay set for the system \eqref{eq:saddle} is shown in gray in fig. \ref{fig:decay_set} where we chose $\mu=-1.1$ and $\mu=0.1$. In fig. \ref{fig:decay_set}(a)-(b) the trajectory at $(x(t), y(t))$ lies in the region where $d(t)$ increases, while in \ref{fig:decay_set}(c)-(d)  the trajectory at $(x(t), y(t))$ lies in the decay set. The initial conditions for this trajectory are the same as those in fig. \ref{fig:unstable_perturb}. The distance from the $z$-axis increases  initially and from about $t= 56$ decreases as shown in fig. \ref{fig:unstable_perturb}(b). This is when the trajectory enters the decay set (shown in gray in fig. \ref{fig:decay_set}(c)). The increase or decrease of the distance of the trajectory from the $z-$axis is determined by whether or not the trajectory lies in the decay set as shown in fig. \ref{fig:decay_set}.

\begin{figure}[!h]

\begin{minipage}{0.48\hsize}
		\begin{center}
			\includegraphics[width =\hsize]{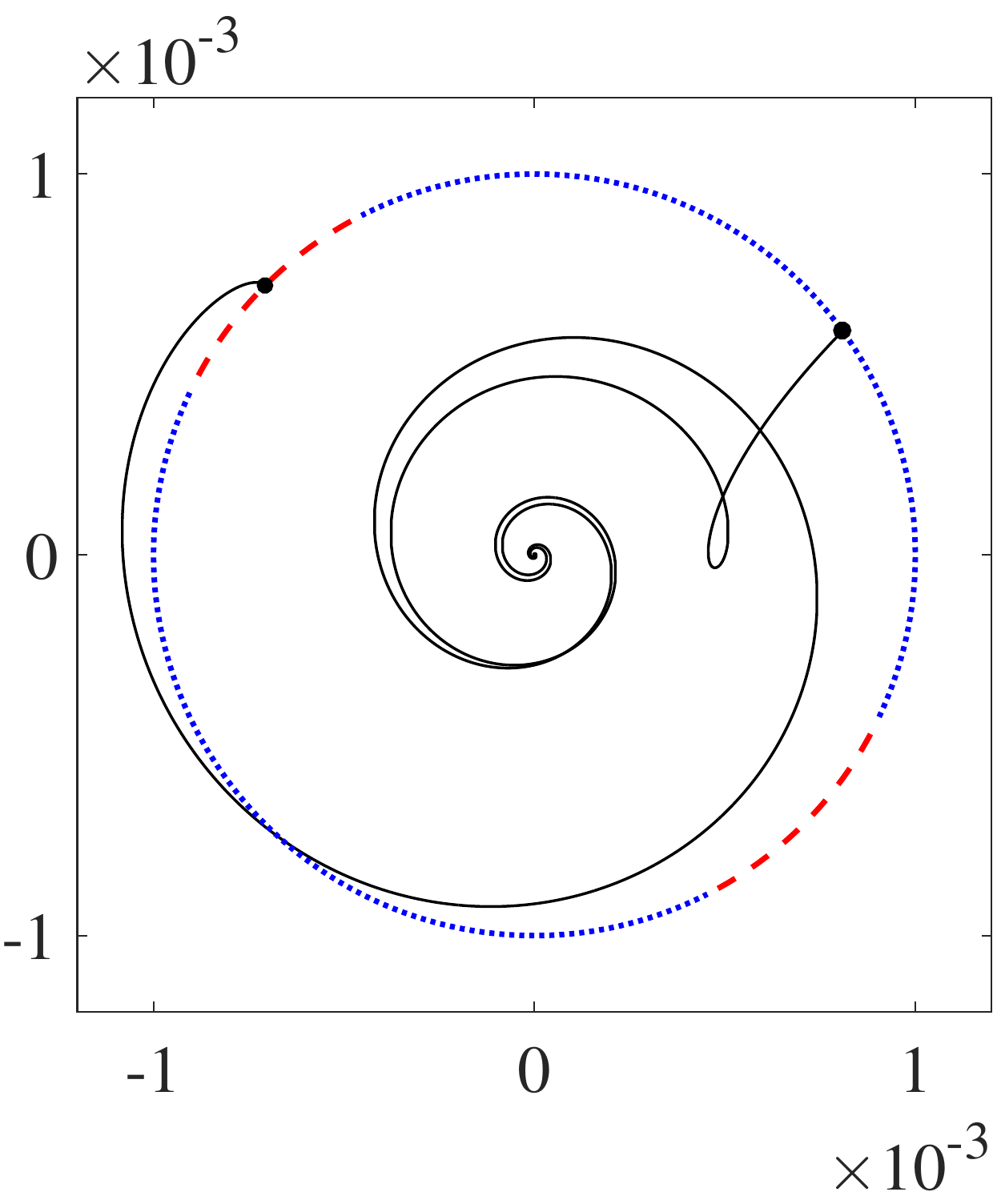}\\
(a)
		\end{center}
	\end{minipage}
	\begin{minipage}{0.48\hsize}
		\begin{center}
			\includegraphics[width =\hsize]{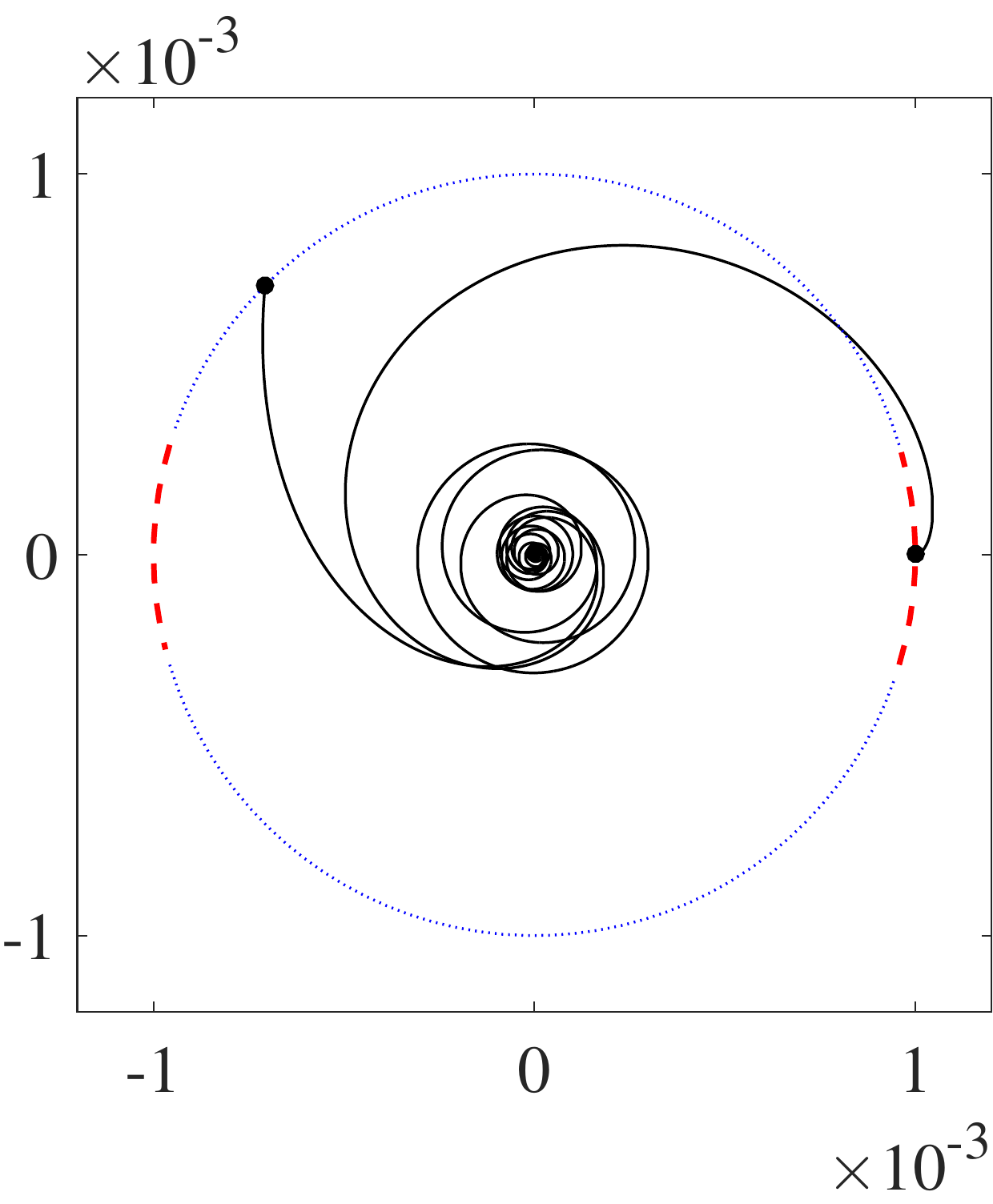}\\
(b)
		\end{center}
	\end{minipage}
\caption{ The variation of the distance of perturbed trajectories. Perturbations are along a circle of radius $10^{-3}$ at (a) $z=\frac{\pi}{4}$ and (b) $z=\frac{\pi}{2}$. The initial perturbation is shown by a black filled circle. Two perturbations are shown in each case. One perturbation is chosen from unstable subset (dashed red circular arc) and one from the stable subset (dotted blue arc).}\label{fig:dt_graph}
\end{figure}

The temporary growth of perturbations and their eventual decay is observed all along the $z-$ axis with the repulsion of  experienced by a trajectory varying. Figure. \ref{fig:dt_graph} shows the projection on to the $x-y$ plane of the evolution of the perturbations at two different values of $z(0)$ away from the origin. A circle whose radius is equal to the norm of the initial perturbation is also shown to illustrate the evolution of the distance of the trajectory from the $z-$axis. The stable and unstable subsets of the circle are shown in blue and red respectively. A sample perturbation from each subset is shown in both the figures. When perturbations begin in the stable subset they first decay while perturbations that begin on the unstable susbet first increase in norm. The local repulsion of a subset of trajectories and the eventual decay of all perturbations is  observed in both cases.

 %Figure \ref{} shows the 

\begin{figure}[!h]

\begin{minipage}{0.48\hsize}
		\begin{center}
			\includegraphics[width =\hsize]{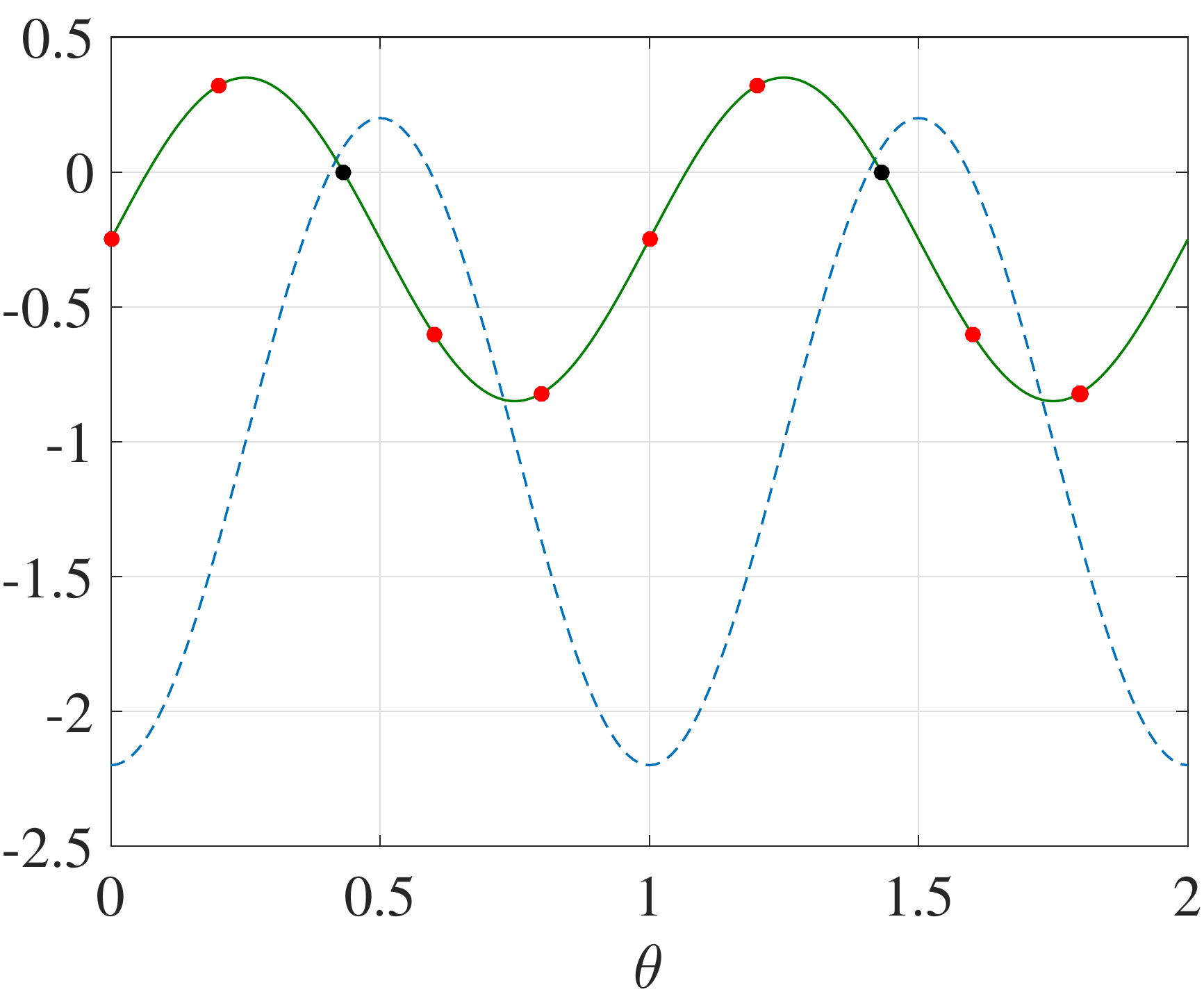}\\
(a)
		\end{center}
	\end{minipage}
	\begin{minipage}{0.48\hsize}
		\begin{center}
			\includegraphics[width =\hsize]{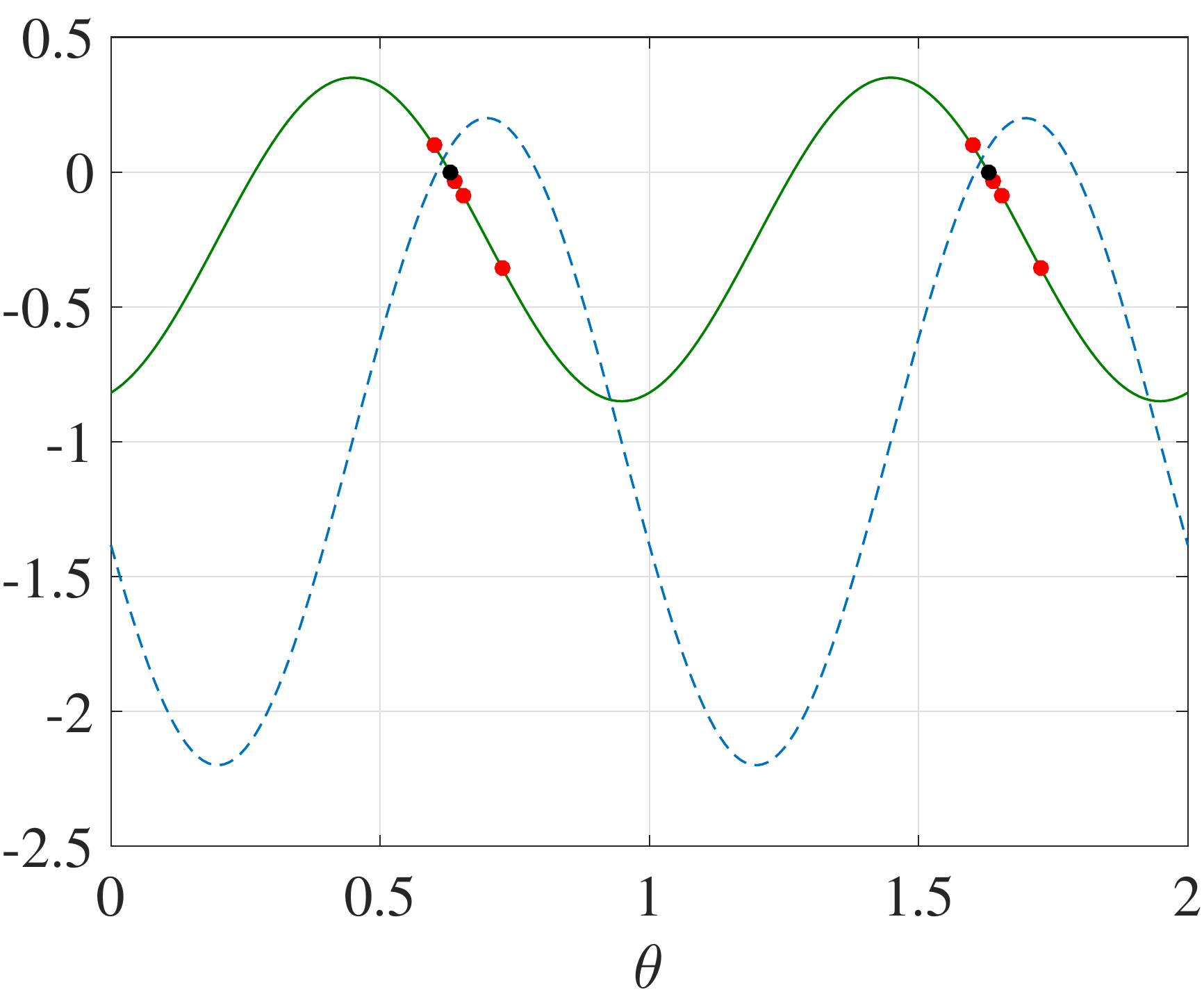}\\
(b)
		\end{center}
	\end{minipage}

\begin{minipage}{0.48\hsize}
		\begin{center}
			\includegraphics[width =\hsize]{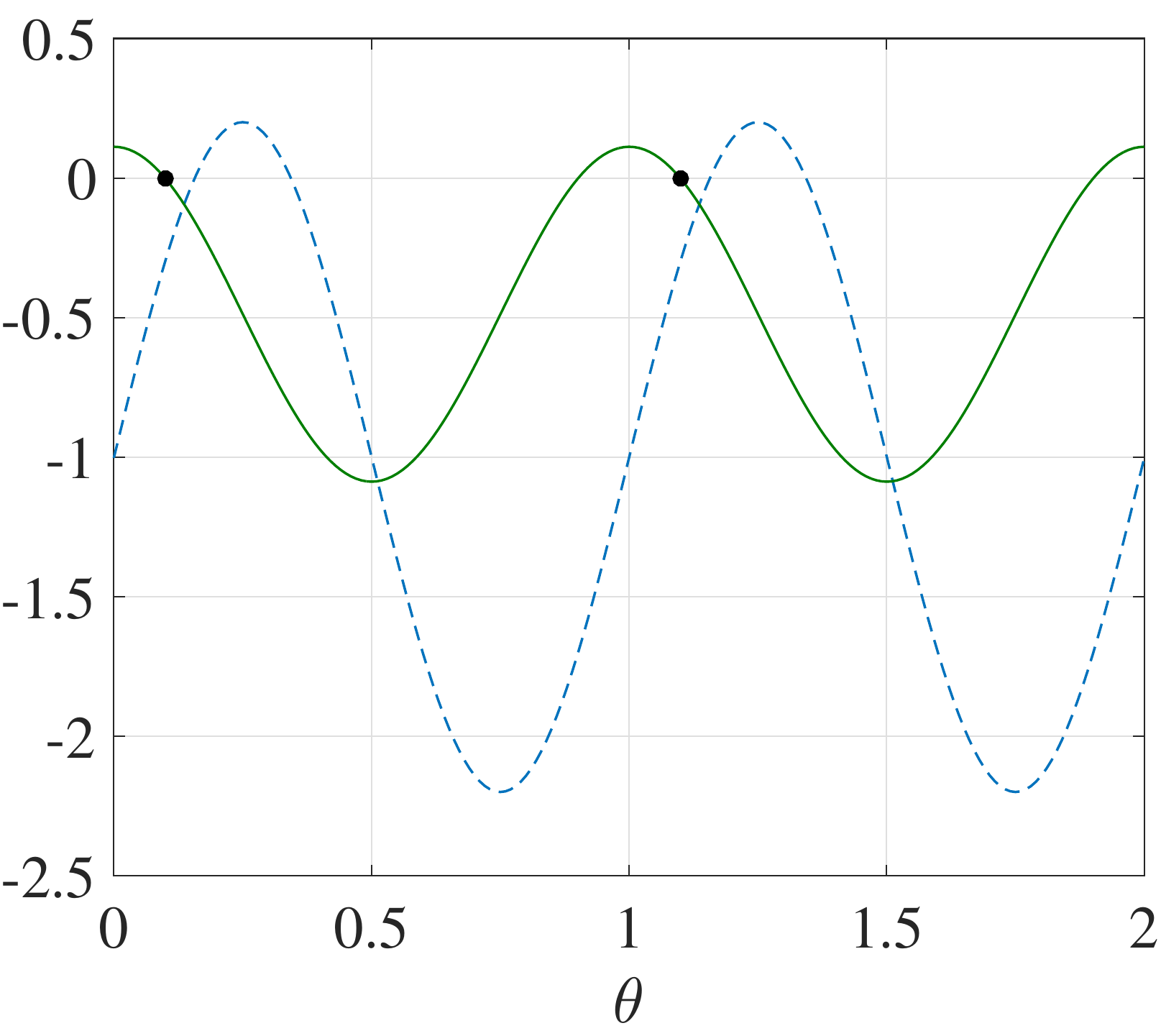}\\
(c)
		\end{center}
	\end{minipage}
	\begin{minipage}{0.48\hsize}
		\begin{center}
			\includegraphics[width =\hsize]{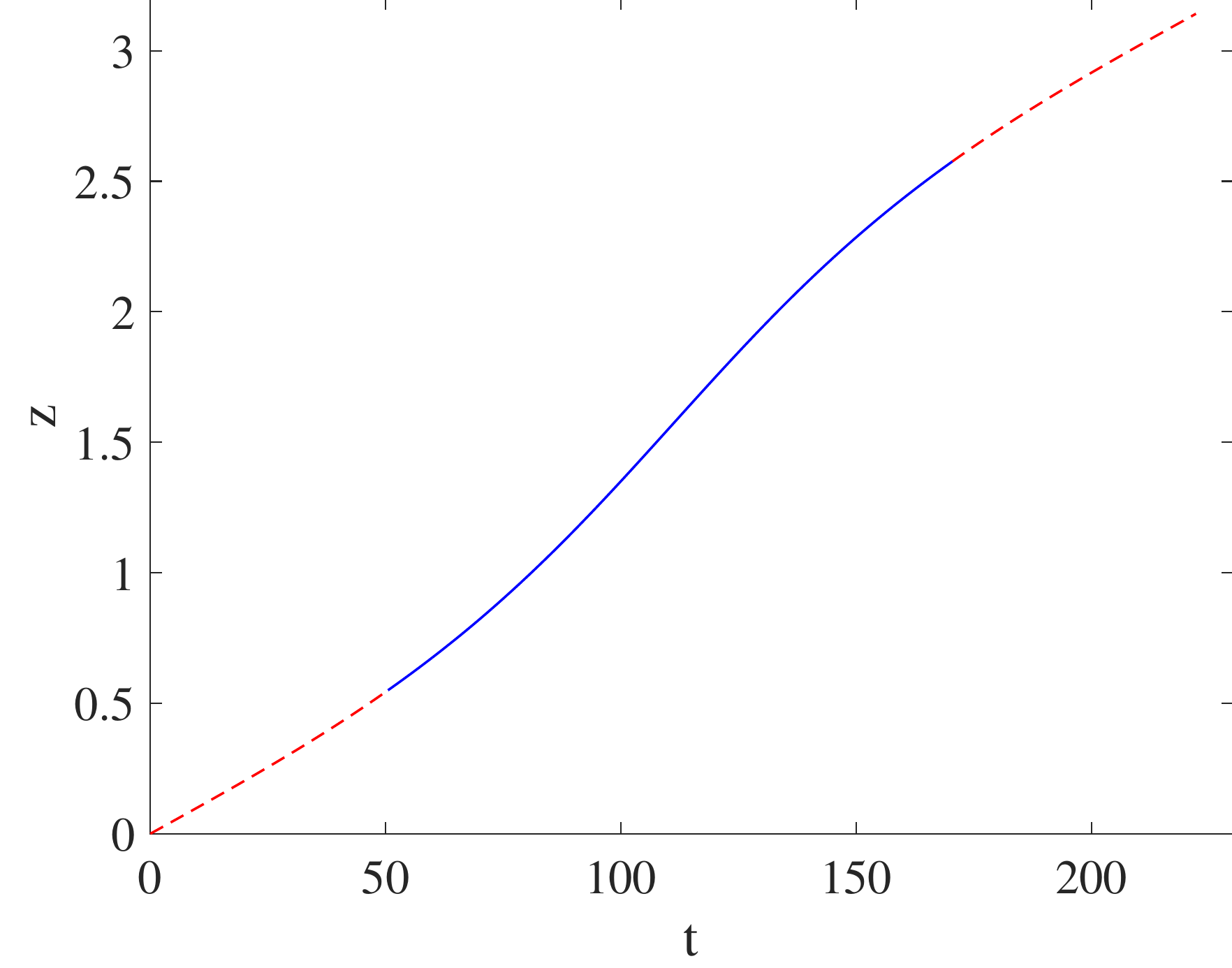}\\
(d)
		\end{center}
	\end{minipage}

\caption{(a) - (c) The green (solid) curve shows $\dot{beta}$ while the blue (dashed) curve shows the instantaneous rate of repulsion/decay ($S$) at (a) $(t,z) = (0,0)$, (b) $(t,z) = (0.25, 0.0025)$ and (c) $(t,z) = (6, 0.0601)$. The stable instantaneous stagnation points of \eqref{eq:beta1} are shown by black filled circles. (a) Eight other perturbations at $z = 0$ with $r=1e-3$ are chosen with initial values of $\theta$ shown by the red filled circles. (b) These perturbations approach towards the stable ISPs and in (c) the eight initial perturbations in $\theta$ almost coincide (with an error of less than $10^-8$) with one of the two stable ISPs. (d) The graph shows $z_{isp}(t)$ with the blue (solid) portion indicating the time period when the ISP lies in the stable set $S^-(z)$ and the red (dashed) portion indicating the time period when the ISP lies in the unstable set $S^-(z)$.  }\label{fig:isp}
\end{figure}

Perturbations are guaranteed to decay only if every trajectory either eventually lies in the stable set or oscillates between the stable and unstable sets but spends more time in the stable set.
Transforming the equation of the dynamical system \eqref{eq:diff_eq} to a polar form, $x = r \cos{\theta}$ and $y = r\sin{\theta}$, where $r = \sqrt{x^2+y^2}$, one can show that
\begin{equation}\label{eq:theta}
\dot{\theta} =\frac{(\lambda_1-\lambda_2)}{2}\sin{(2\omega z-2\theta)}.
\end{equation}
We define a new variable $\beta = \theta - \omega z$, which represents the relative polar angle of the trajectory with respect to the stable and unstable eigenvectors at some $z$. Then
\begin{align}\label{eq:beta1}
\dot{\beta} &= \dot{\theta}- \omega \dot{z} = -\frac{(\lambda_1-\lambda_2)}{2}\sin{(2\beta)} - \omega \dot{z} \nonumber \\
&= -\frac{(\lambda_1-\lambda_2)}{2}\sin{(2\beta)} - \omega k (1+\sin^2{z}).
\end{align}
Equation \eqref{eq:beta1} is independent of $r$, but depends on $z$. When $\omega k$ is small instantaneous stagnation points (ISP) of \eqref{eq:beta1} plays a dominant role in the evolution of the $\beta$. At each $z(t)$, the ISPs are obtained by setting the right hand side of \eqref{eq:beta1} to zero. There are four ISPs of \eqref{eq:beta1}. For example at $z=0$, fig. \ref{fig:isp}(a) shows $\dot{\beta}$ by the solid green curve which is zero at four distinct values of $\theta$. Two of these, shown by black solid dots, are stable due to the negative slope. The time evolution of these ISPs are the so called distinguished hyperbolic trajectories and all solutions of \eqref{eq:beta1} converge to a neighborhood of these distinguished trajectories, \cite{wiggins_npg_2002}.  Figure \ref{fig:isp} (a) - (c) shows such a convergence of eight solutions of \eqref{eq:beta1} with distinct values of $\theta(0)$ shown by the red filled circles. The growth or decay of perturbations of the dynamical system \eqref{eq:diff_eq}  to $(x,y) = (0,0)$  is determined by the evolution of the stable ISPs of \eqref{eq:beta1}.  As the position along $M$ varies, the two stable ISPs could lie either in the stable set $S^-(z)$or the unstable set $S^+(z)$. The blue (dashed) curve in fig. \ref{fig:isp} (a)-(c) shows the instantaneous rate of repulsion, $S$. When the stable ISP lies in the stable set (where the graph of $S$ is negative) a perturbation away from the $z-$ axis decays.  Numerical simulations show that the stable ISPs  lie in the stable set $S^-(z)$ for a longer duration of time than in the set $S^+(z)$, \ref{fig:isp}(d). Since $\dot{z}$ is periodic, with period $\pi$, the graph of $z_{isp}(t)$ in \ref{fig:isp}(d) is also periodic. The repulsion or decay rate of a trajectory is repeated along the $z-$ axis. Furthermore the highest instantaneous rate of repulsion of perturbations is smaller than the highest instantaneous rate of attraction if $\lambda_1+\lambda_2 <0$, see for example fig. \ref{fig:S}. This leads to the eventual decay of all perturbations.

Figure \ref{fig:isp} also explains the weaker repulsion of perturbations seen in fig. \ref{fig:dt_graph} compared to the case shown in fig. \ref{fig:unstable_perturb}. The observed repulsion is the least in the neighborhood of $z = (2n+1)\frac{\pi}{2}$ for any integer $n$. This is because the stable ISPs lie in the stable set $S^-$ in a large neighborhood of $z=\frac{\pi}{2}$, fig. \ref{fig:isp}(d). Since convergence of any solution of \eqref{eq:beta1} to one of the stable ISPs occurs very rapidly, any perturbation whether it begins in the stable or the unstable set, decays rapidly, experiencing only a small transient repulsion. Repulsion of perturbations around the $z$ axis is the highest in the neighborhood of $z = n\pi$, since the stable ISPs lie in the unstable set $S^+(z)$ in a large neighborhood of $z = n\pi$, see fig. \ref{fig:isp}(d).

When $\omega k \gg \frac{\lambda_1 - \lambda_2}{2}$, instantaneous stagnation points do not exist for some or all values of $z$ and the relative angle $\beta$ is driven to oscillate with a large amplitude $(\approx \omega k)$. In this case a trajectory, $(x(t), y(t), z(t))$ moves into and out of the decay set $S^-(z)$ almost periodically. However since the rate of decay in the decay set is larger than the rate of repulsion, the perturbation eventually decays with oscillations. An example of such a case is shown in fig. \ref{fig:w_98}(b), where the norm of a perturbation decays to zero with oscillations.

\subsection{Parametric dependence of global stability of the invariant manifold}

It should be obvious that if $\omega =0$ then the $z-$axis is both locally and globally unstable. Any normal perturbation such that $y(0) \neq 0$ would lead to the trajectory diverging from the $z-$ axis. Therefore the global stability of the $z-$ axis certainly depends on the rate of the rotation of the stable and unstable eigenvectors along the $z-$ axis. Starting from $\omega=0$, we performed numerical simulations for increasing values of $\omega$ to determine the critical value of the rate of rotation at which the $z-$ axis become globally stable. These simulations show that the transition of the $z$-axis to a globally stable manifold occurs in four stages. In each of these stages the behavior of the function $d(t)$ the distance of a trajectory from the $z$-axis is qualitatively distinct. 

\begin{figure}[!h]

\begin{minipage}{0.48\hsize}
		\begin{center}
			\includegraphics[width =\hsize]{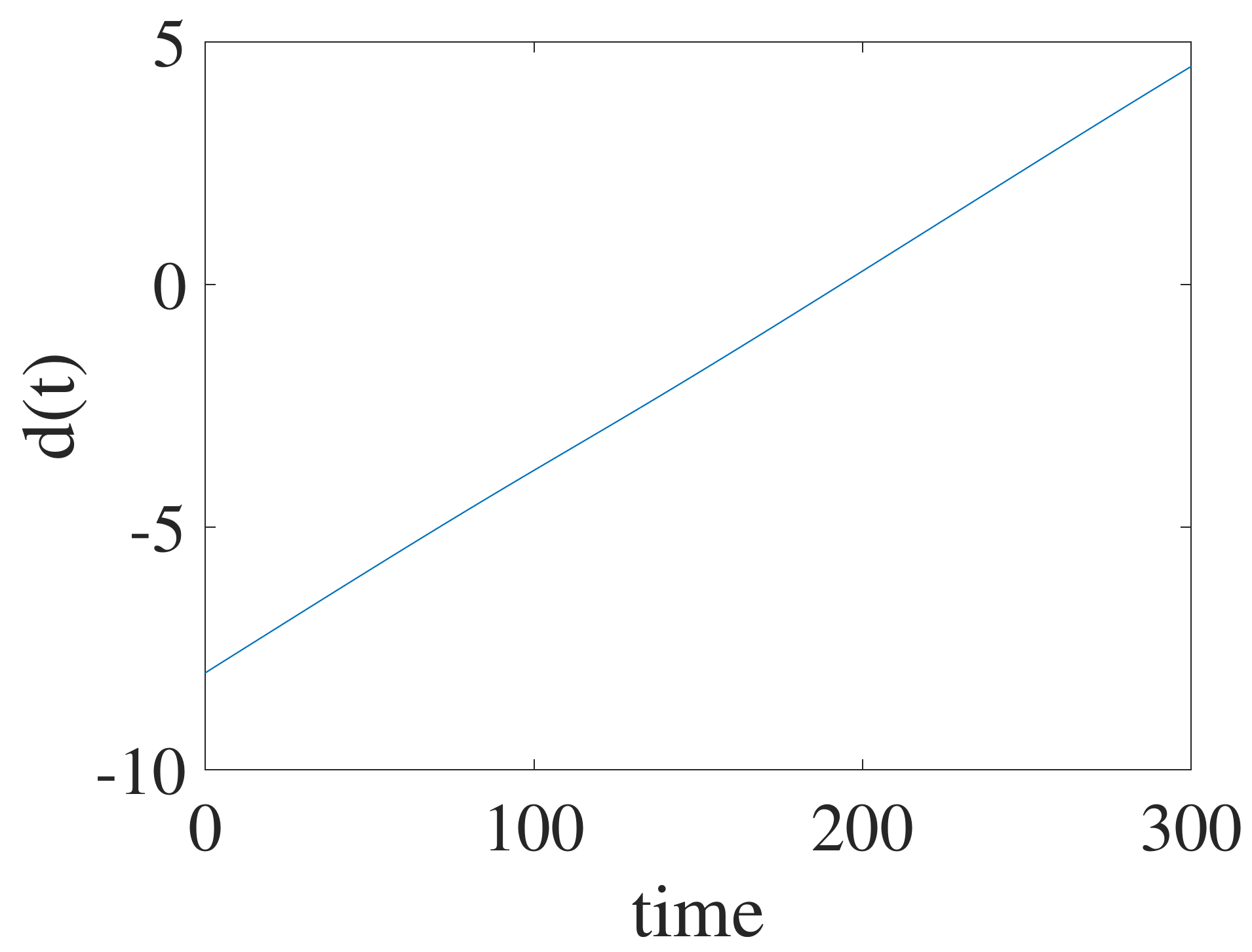}\\
(a)
		\end{center}
	\end{minipage}
	\begin{minipage}{0.48\hsize}
		\begin{center}
			\includegraphics[width =\hsize]{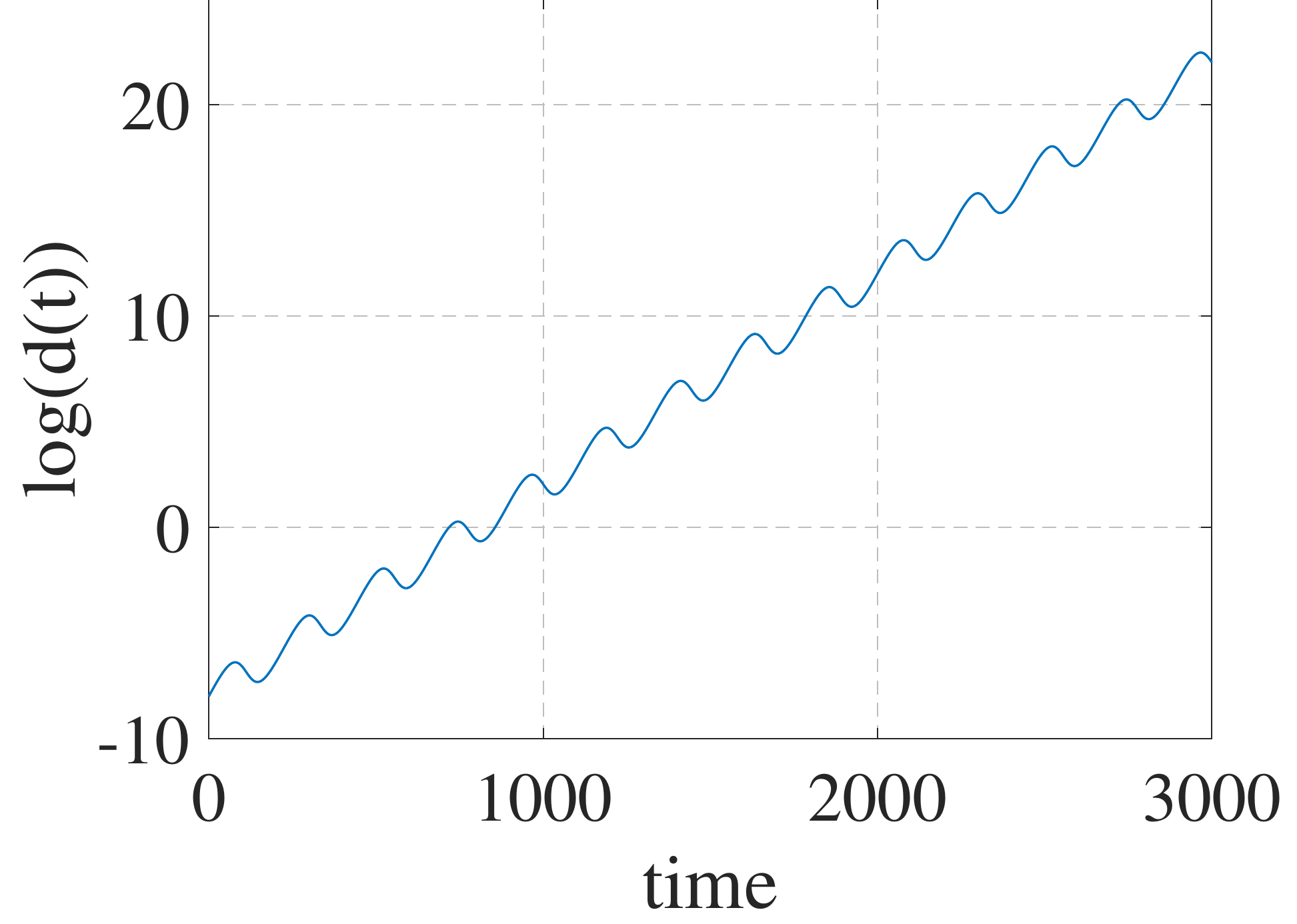}\\
(b)
		\end{center}
	\end{minipage}

\begin{minipage}{0.48\hsize}
		\begin{center}
			\includegraphics[width =\hsize]{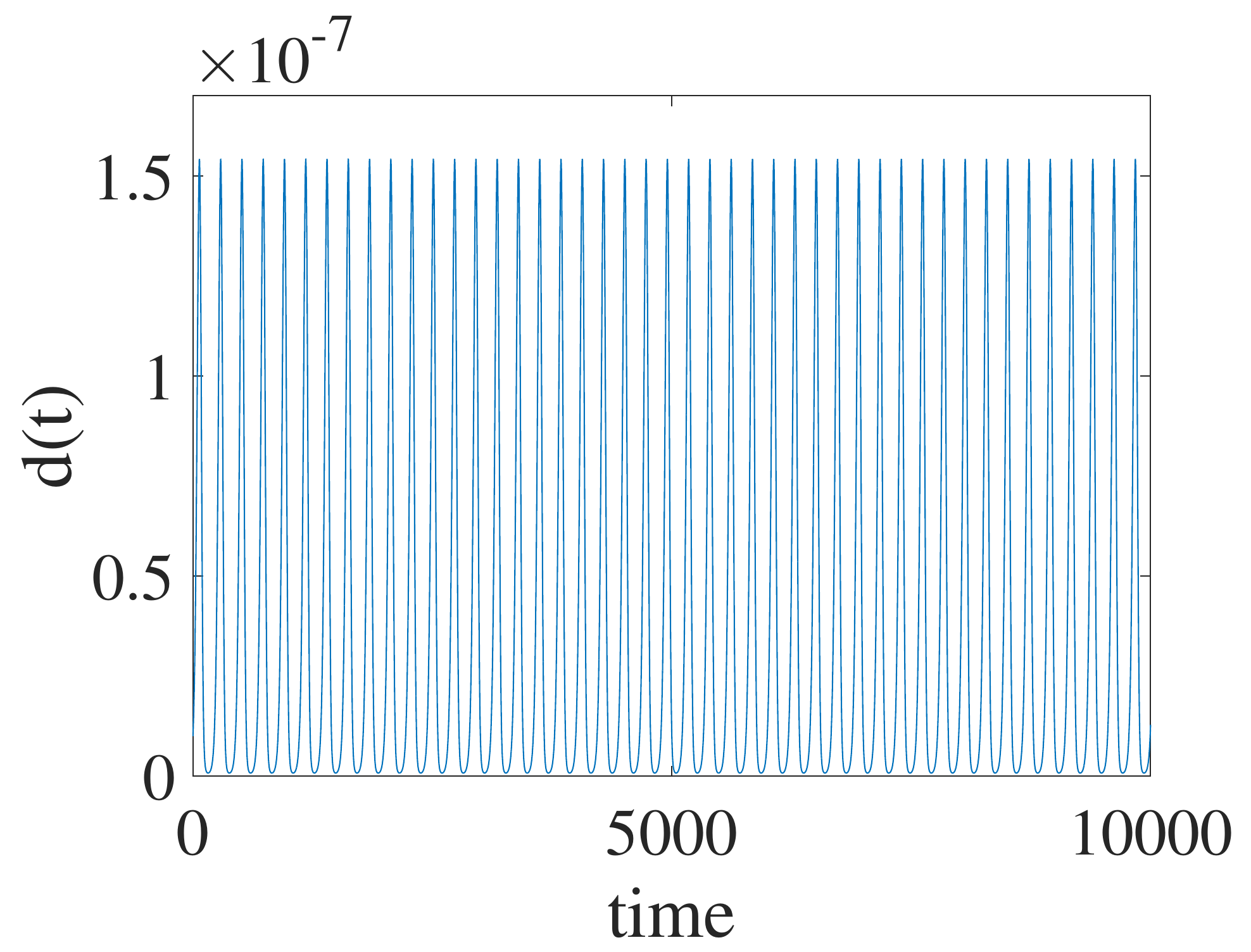}\\
(c)
		\end{center}
	\end{minipage}
	\begin{minipage}{0.48\hsize}
		\begin{center}
			\includegraphics[width =\hsize]{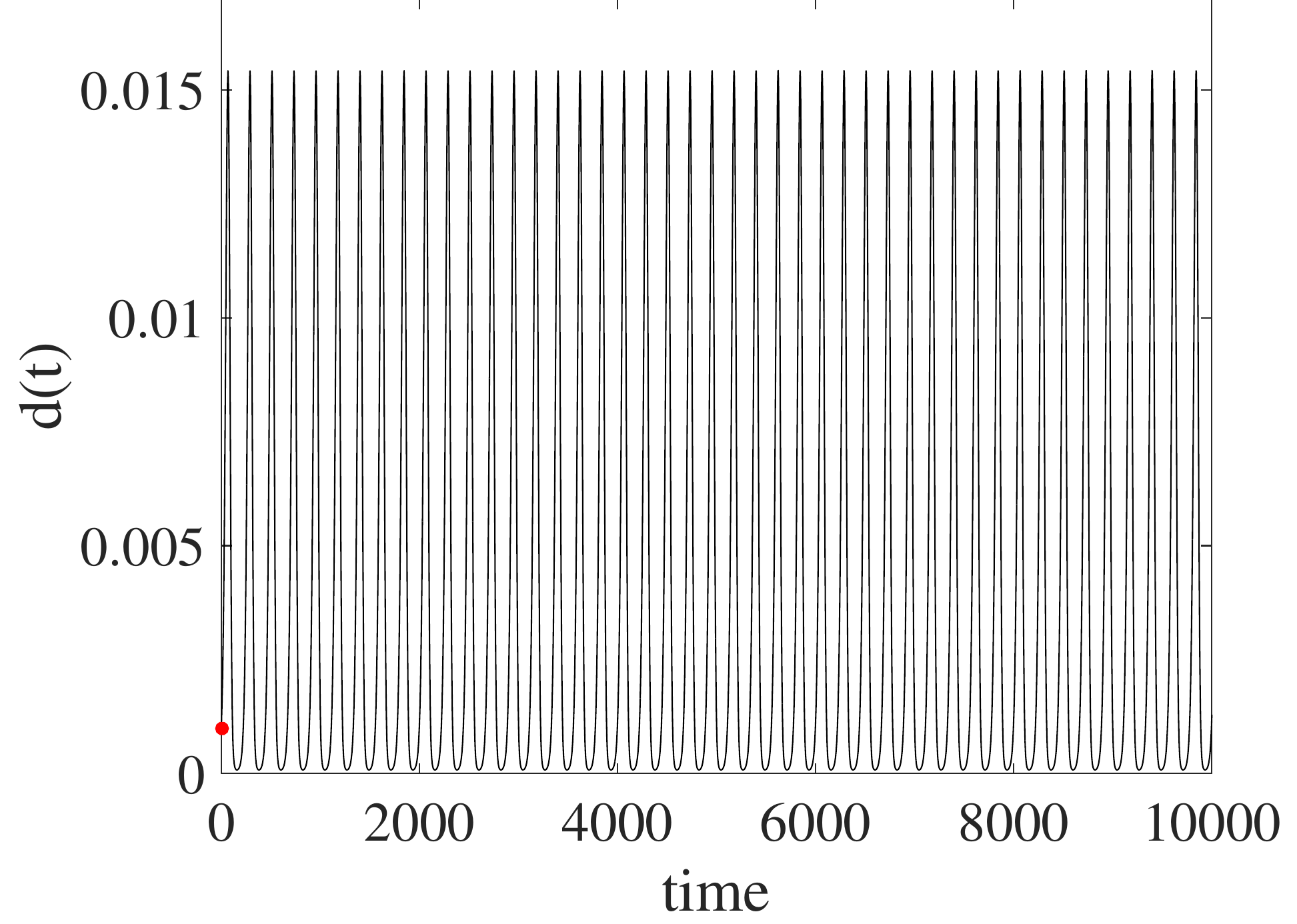}\\
(d)
		\end{center}
	\end{minipage}
\caption{ Distance, $d(t)$ (or $log_{10}d(t)$) of the trajectory from the $z-$ axis with initial conditions $(0, 10^{-7})$. (a) At $\omega = 5$, the distance grows exponentially. (b) At $\omega = 20$, the distance grows but shows oscillations.  (c) At $\omega = 22.4686859$, the distance is bounded and oscillatory for perturbations of any norm, for example $ d(0) = 10^-{8}$ and (d) $d(0) = 10^{-3}$. }\label{fig:stages_z}
\end{figure}

\begin{figure}[!h]

\begin{minipage}{0.48\hsize}
		\begin{center}
			\includegraphics[width =\hsize]{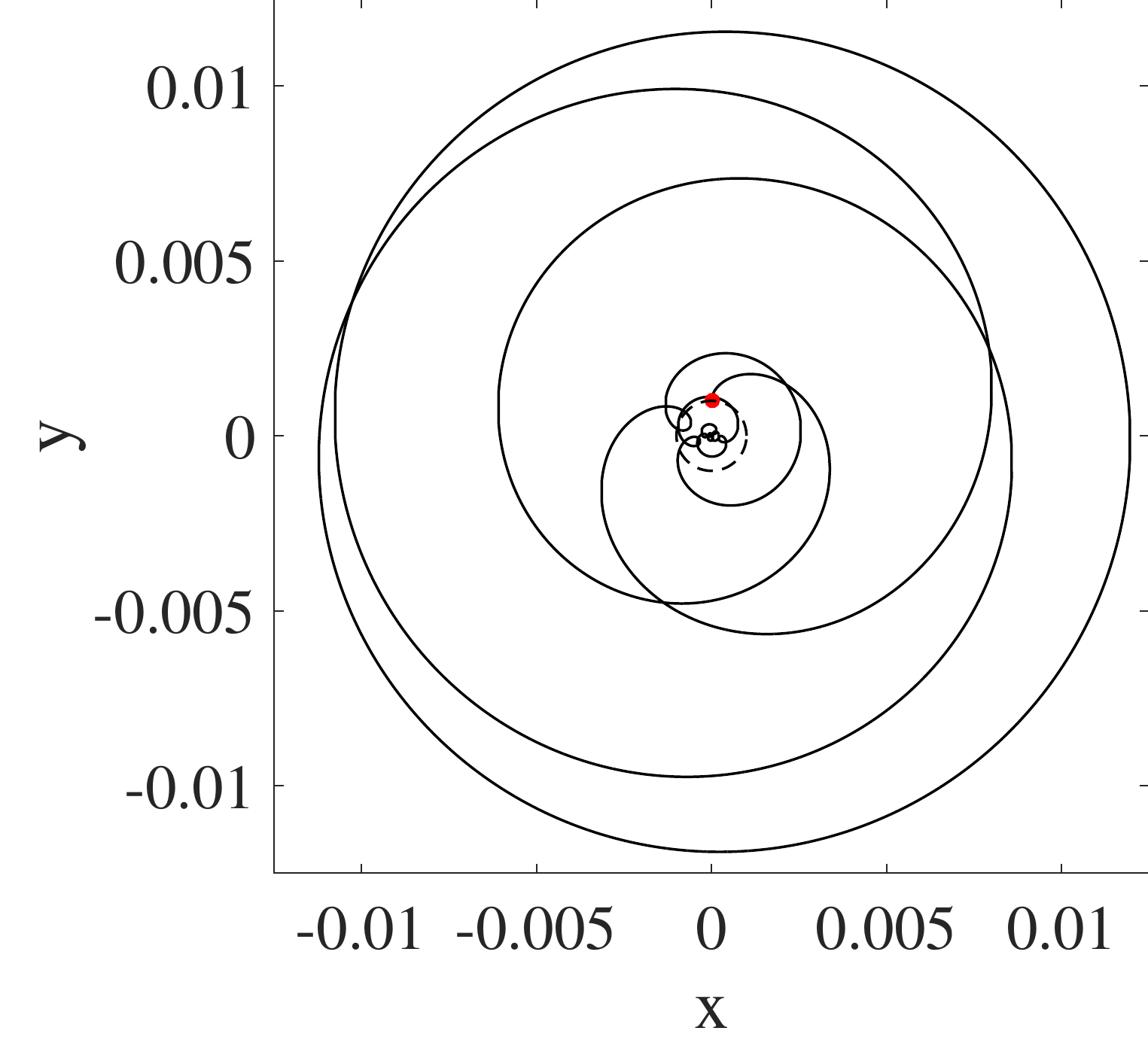}\\
(a)
		\end{center}
	\end{minipage}
	\begin{minipage}{0.48\hsize}
		\begin{center}
			\includegraphics[width =\hsize]{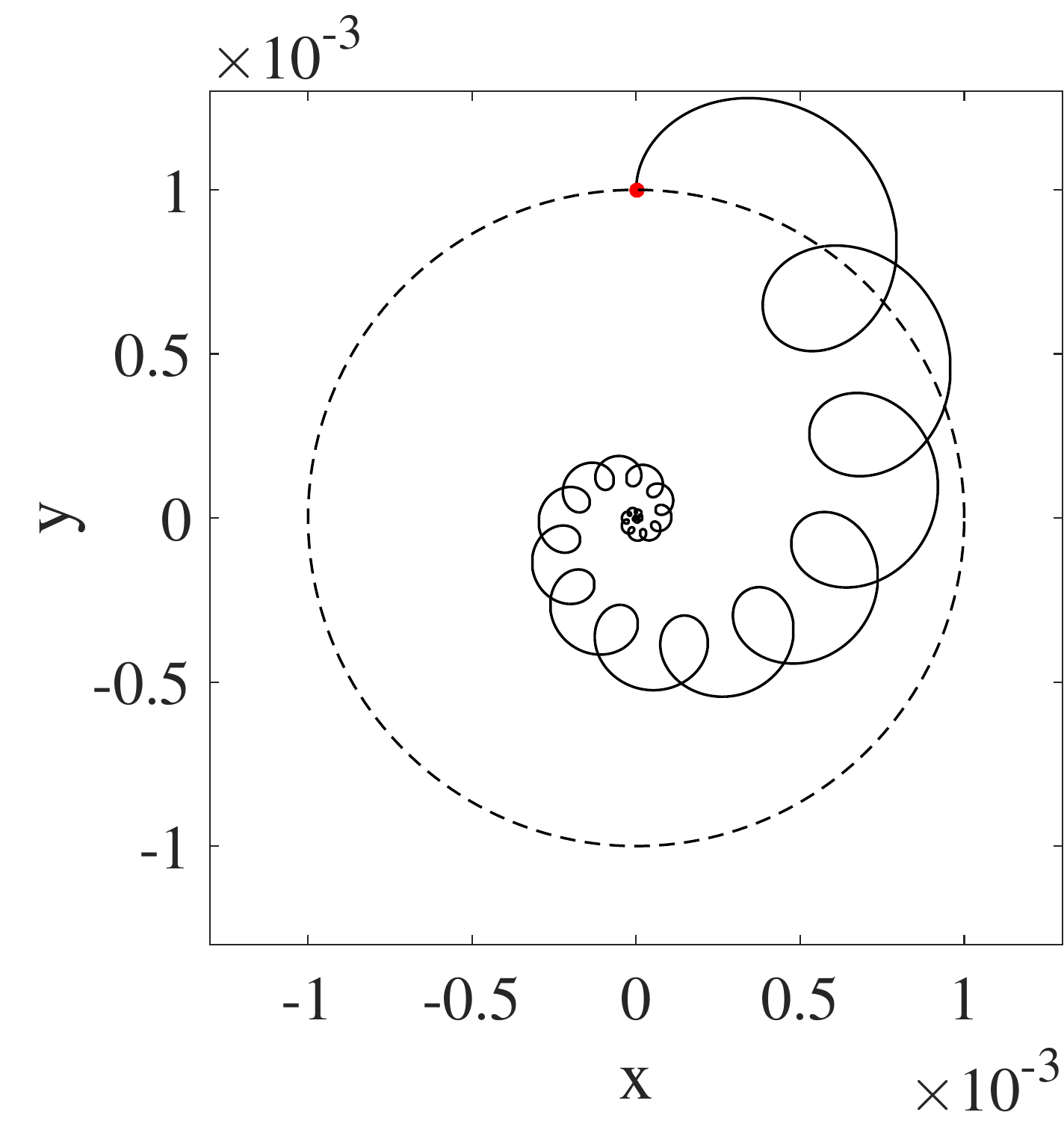}\\
(b)
		\end{center}
	\end{minipage}

\caption{ Distance, $d(t)$ of the trajectory from the $z-$ axis when $\lambda_1 = -1.1$, $\lambda_2 = 0.9$ and $\omega = 98$. The initial conditions are (a) $(x,y,z) = (0, 10^-{3},0)$ and (b) $(x,y,z) = (0, 10^-{3},\frac{\pi}{2})$. In (a) the distance grows more than 10 times its initial value and in (b) it grows about $1.4$ times its initial value before decaying to zero.  }\label{fig:w_98}
\end{figure}

For  small values of $\omega$ any trajectory diverges away from $z-$ axis and $d(t)$ is a monotonically increasing function, see fig. \ref{fig:stages_z}(a). In the second stage the distance of a trajectory undergoes small oscillations with the mean distance increasing. Figure \ref{fig:stages_z}(b) shows these oscillations in the $log_{10}(d(t))$. We identified the local maxima of the function $log_{10}(d(t))$ and find that these maxima are linearly increasing. As $\omega$ increases further, the growth in these local maxima of   $log_{10}(d(t))$  decreases. At a certain critical value of the rate of rotation, $\omega_{cr} \approx  22.4686859$, we find that the distance function, $d(z)$, is periodic, see fig. \ref{fig:stages_z}(c).  Numerical integration of \eqref{eq:diff_eq} for very large periods of time yields a trajectory that is bounded for perturbations across a range of magnitudes. Figures  \ref{fig:stages_z}(c) and \ref{fig:stages_z}(d) show the periodic variation in the distance of a trajectory from the $z-$axis when the norm of the initial perturbation is $10^{-8}$ and $10^{-3}$ respectively. The time period between consecutive maxima in the graphs is $T = 222.1442$  accurate up to three decimal places. Identifying the local maxima of the distance of this trajectory, shows that these maxima are nearly constant with variation on the order of $10^{-13}$ and $10^{-9}$. The numerics therefore indicate bounded trajectories whose distance from the $z-$axis varies periodically. For $\omega > \omega_{cr}$ normal perturbations from the $z-$ axis decay with the distance decreasing with oscillations. Increasing the value of $\omega$ even up to $20000$ showed that all normal perturbations do not decay monotonically. This is to be expected since any perturbation in the locally repelling direction has to first increase before decaying. 
\begin{figure}[!h]
		\begin{center}
			\includegraphics[width =0.9\hsize]{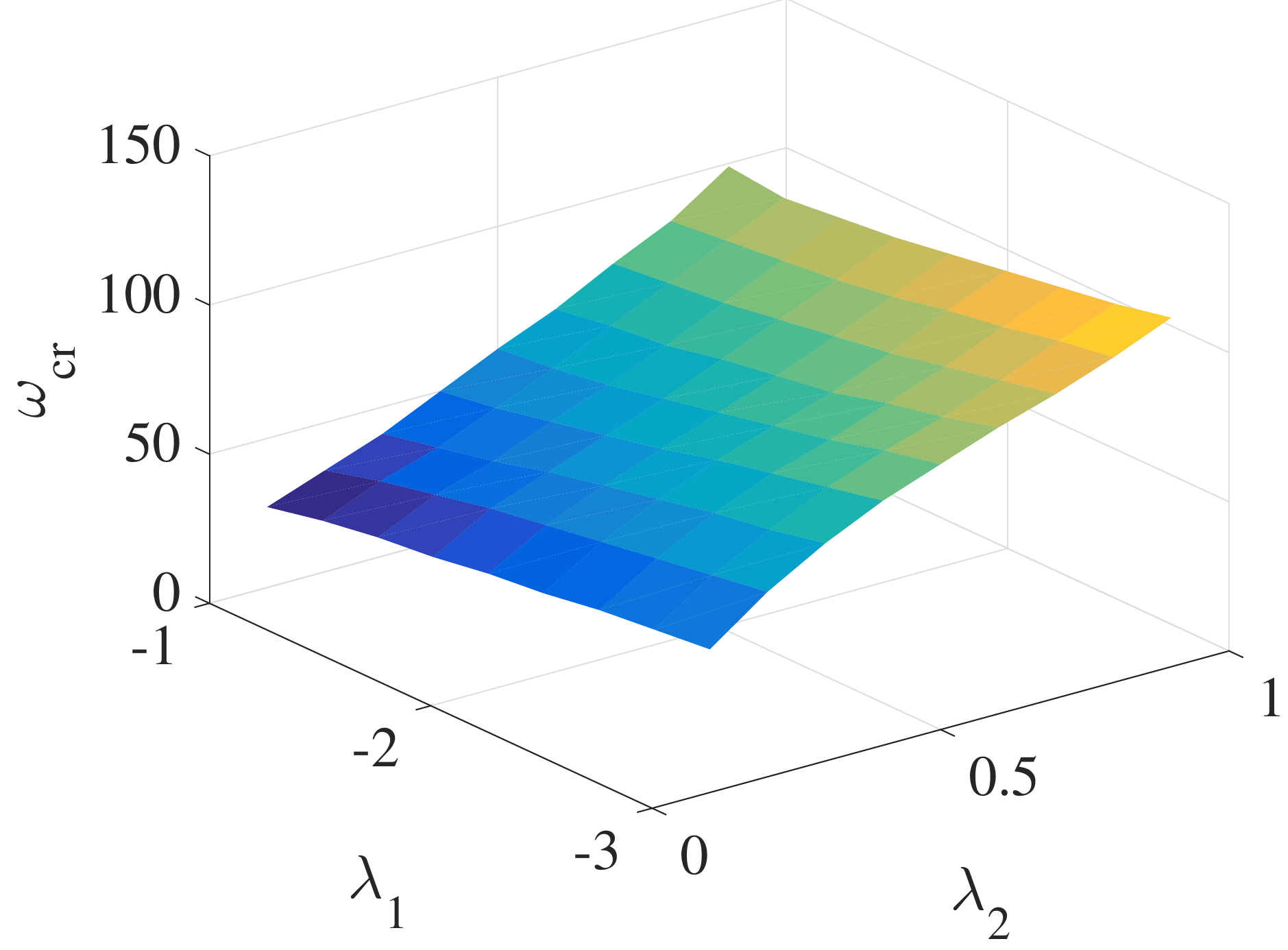}
		\end{center}
\caption{The value of $\omega_{cr}$ increases with both $|\lambda_1|$ and $\lambda_2$.}\label{fig:wcr}
\end{figure}
The value of $\omega_{cr}$ depends on $\lambda_1$ and $\lambda_2$ for a fixed vector field $\dot{z}$. The critical angular velocity, $\omega_{cr}$, increases with $\lambda_2$ and the magnitude of $\lambda_1$. This relationship is shown in fig. \ref{fig:wcr}. When $\lambda_1+\lambda_2$ is closer to zero, the temporary repulsion of the trajectories from the $z-$axis is more dramatic, as shown in fig. \ref{fig:w_98}. The $z-$ axis loses its global stability if $\lambda_1 + \lambda_2>0$. 

\section{A gobally stable limit cycle that is locally unstable everywhere} \label{sec:limit_cycle}
The invariant manifold of the dynamical system \eqref{eq:sys} is unbounded. We construct a dynamical system in $\mathbb{R}^3$ that has an invariant limit cycle. The equations for this dynamical system are arrived by forcing the normal stable and unstable eigenvectors at each point on the limit cycle to undergo a rotation. We express these equations in cylindrical coordinates, where the rotation of the stable and unstable eigenvectors is easily seen. The equations are
\begin{align}\label{eq:sys2}
\begin{pmatrix}
\dot{r}\\
\dot{y}
\end{pmatrix} &= 
\textbf{A}(\theta)\begin{pmatrix}
r-1\\
y
\end{pmatrix} \nonumber \\
\dot{\theta} &= 0.01(1+\sin^2{\theta})
\end{align}
where 
\begin{equation} \label{eq:A2}
\textbf{A}(\theta) = \textbf{R}\begin{pmatrix}
\lambda_1 & 0\\
0 & \lambda_2
\end{pmatrix} \textbf{R}^{-1}.
\end{equation} 

As in the earlier example, we choose $\lambda_1<0$ and $\lambda_2>0$. The set
\begin{equation}\label{eq:lim_cycle}
M = \{(r,y,\theta)|r=1, y= 0\}
\end{equation}
is an invariant manifold (a limit cycle) for the system \eqref{eq:sys2}. The limit cycle repels almost all normal perturbation for an intermediate period of time. If the rate of rotation, $\omega$ is sufficiently high, these  repelled trajectories eventually converge to the limit cycle. 

\begin{figure}[!h]

	\begin{minipage}{0.48\hsize}
		\begin{center}
			\includegraphics[width =\hsize]{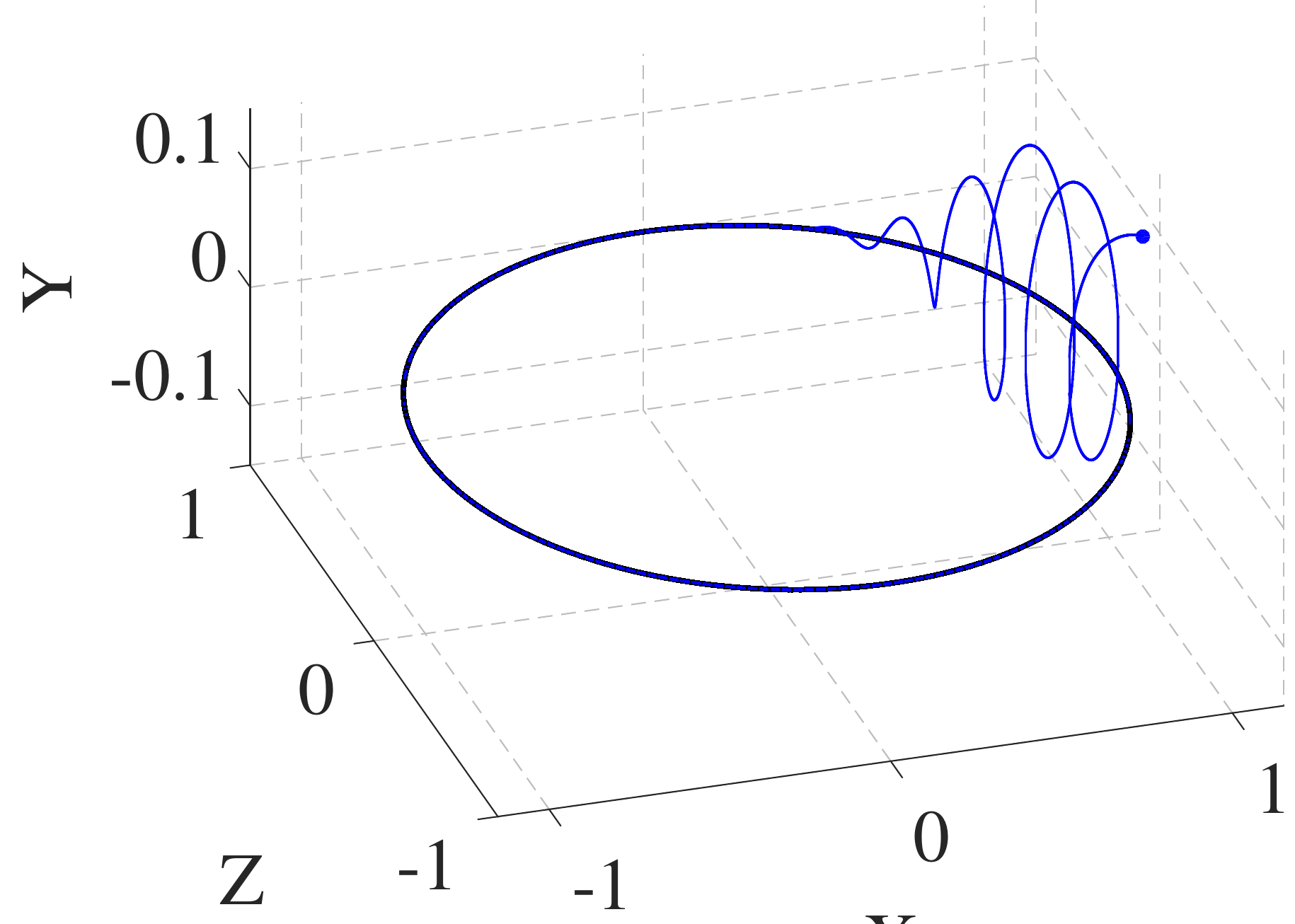}\\
(a)
		\end{center}
	\end{minipage}
\begin{minipage}{0.48\hsize}
		\begin{center}
			\includegraphics[width =\hsize]{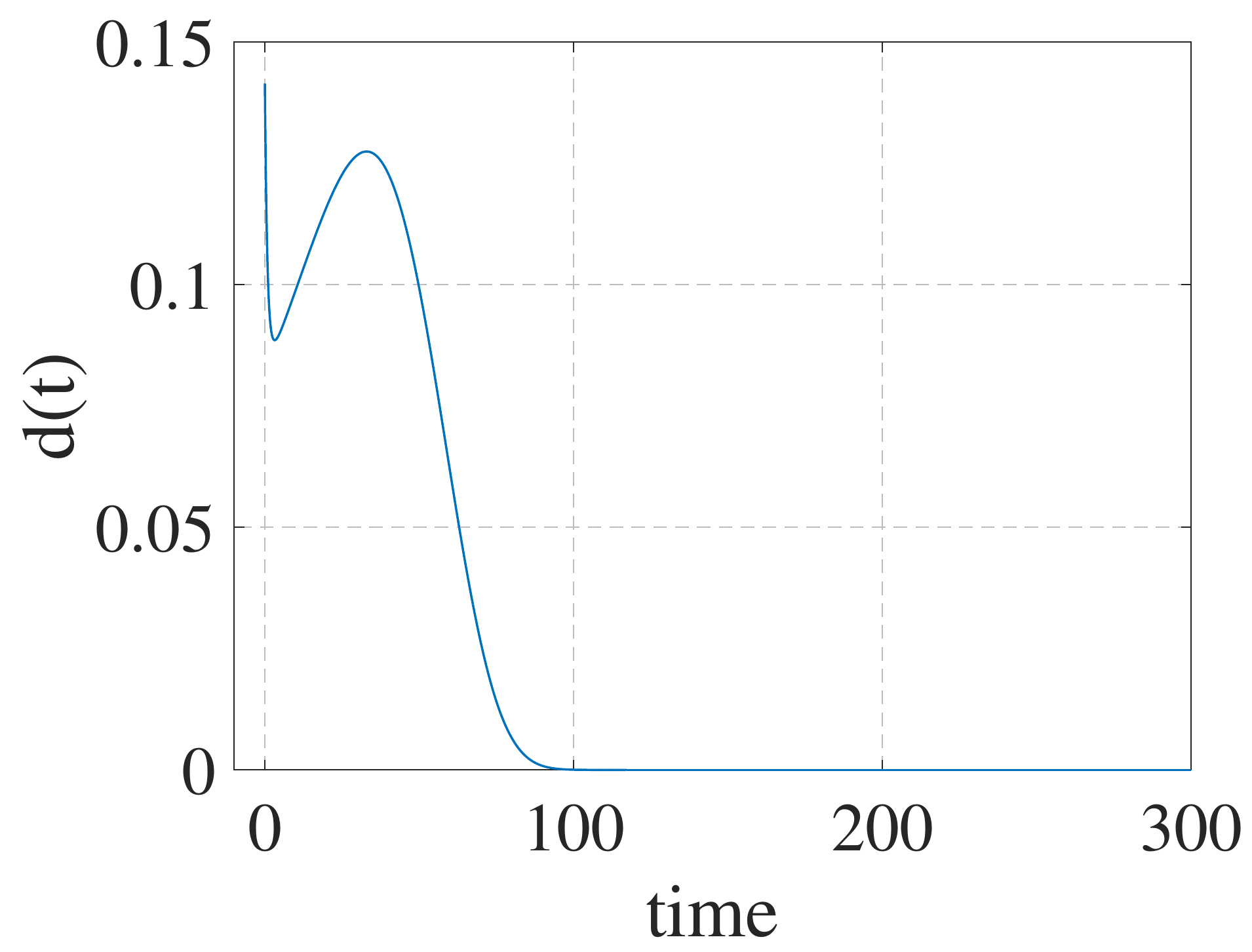}\\
(b)
		\end{center}
	\end{minipage}

	\begin{minipage}{0.48\hsize}
		\begin{center}
			\includegraphics[width =\hsize]{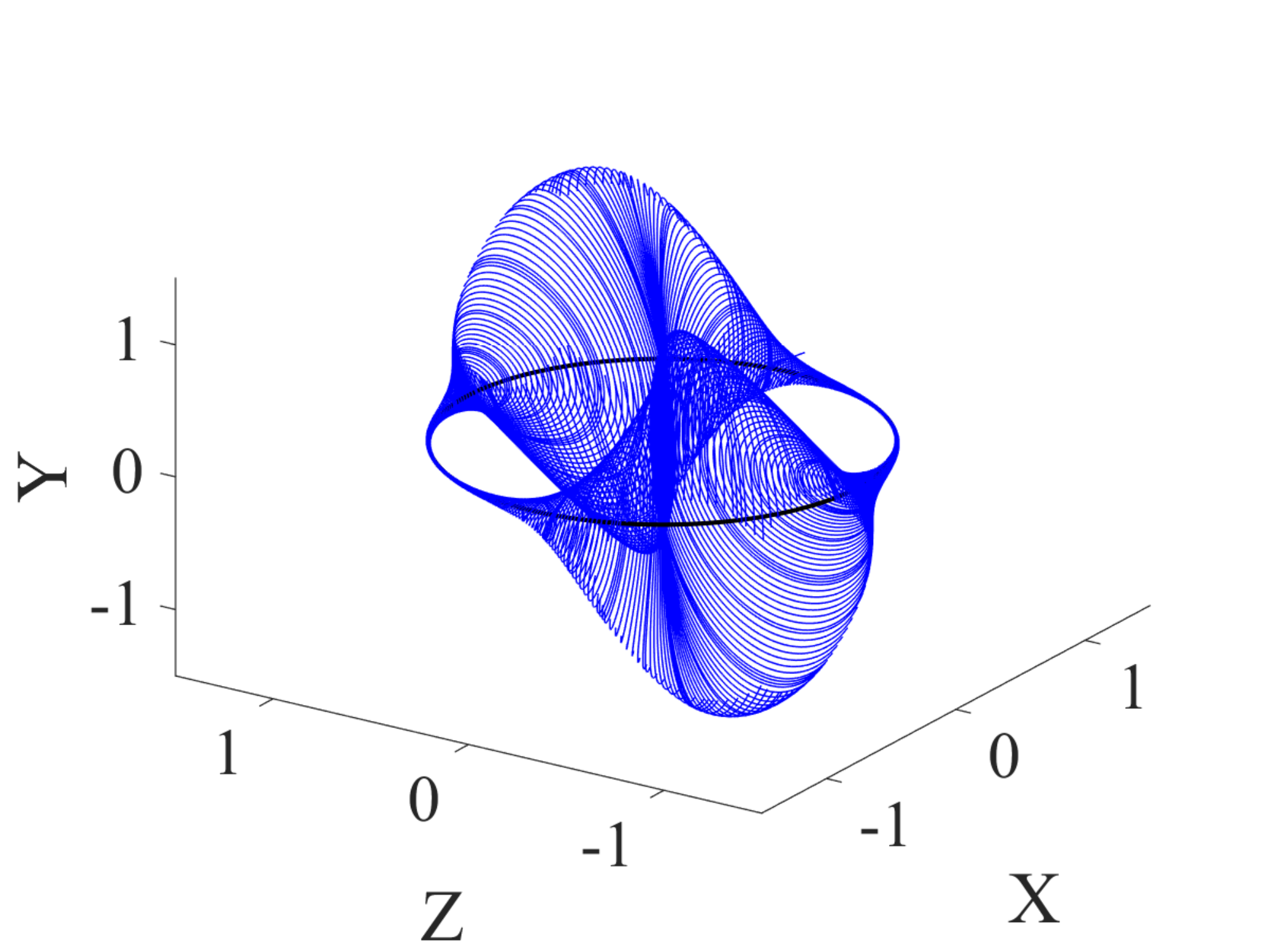}\\
(c) 
		\end{center}
	\end{minipage}
\begin{minipage}{0.48\hsize}
		\begin{center}
			\includegraphics[width =\hsize]{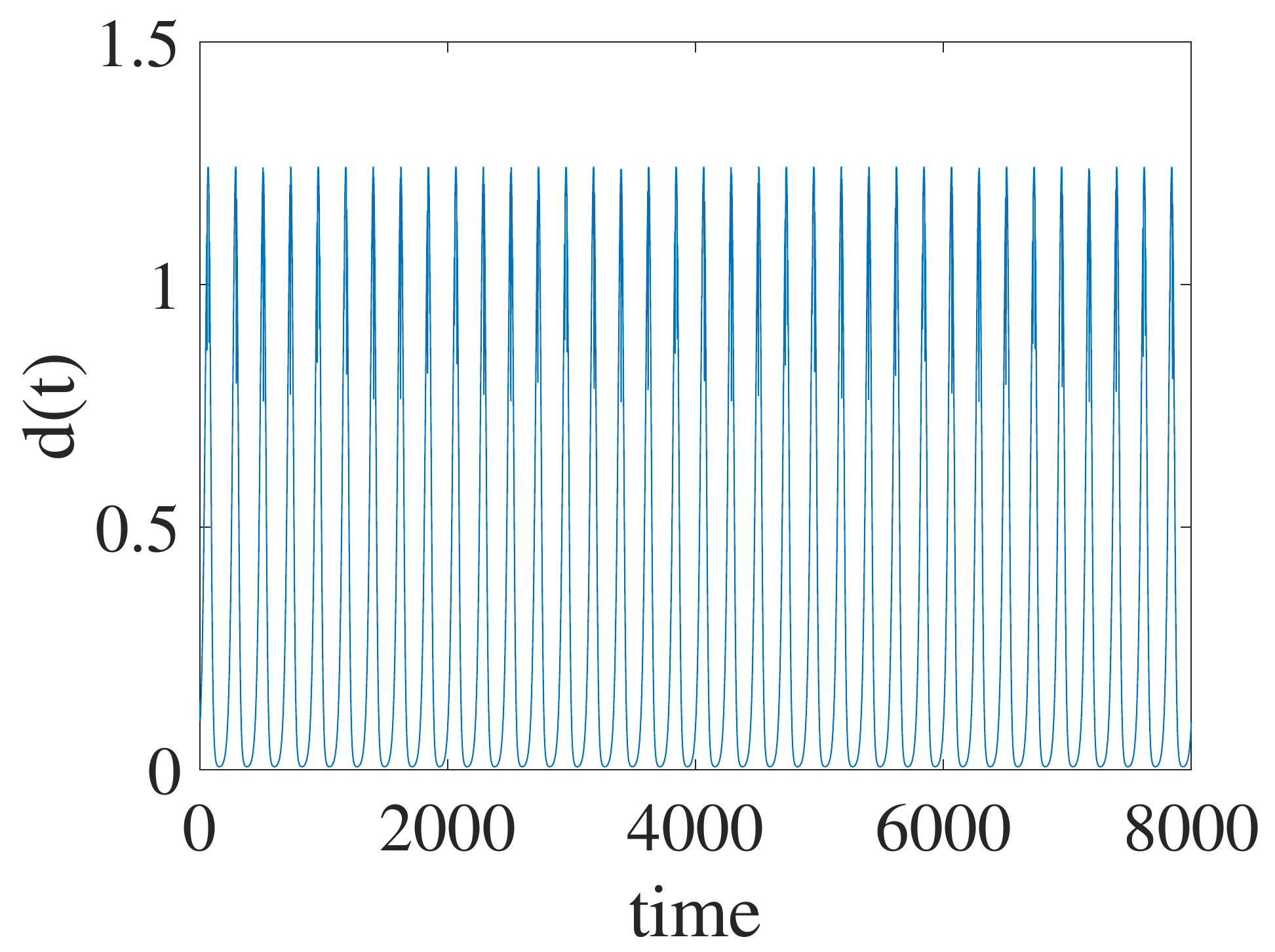}\\
(d) 
		\end{center}
	\end{minipage}

\caption{ (a) A typical trajectory for the system \eqref{eq:sys2}. The systems parameters are $\lambda_1 = -1.1$, $\lambda_2 = 0.1$, $\omega = 30$. The initial conditions for the trajectory are $r(0) = 1.1$ and $y(0) = 0.1$. (b) The distance, $d(t)$ of the trajectory from the limit cycle, eventually converges to zero.  (c) A typical trajectory for the system \eqref{eq:sys2} when $\omega = \omega_{cr}$. The trajectory is bounded but does not converge to the limit cycle. (d) The distance, $d(t)$ of the trajectory from the limit cycle, is a periodic function.}\label{fig:traj_limit_cycle} 
\end{figure}

Figure \ref{fig:traj_limit_cycle}(a) shows a trajectory with initial conditions $(r(0) = 1.1, y(0) = 0.1, \theta =0)$. The trajectory spirals around the limit cycle and converges to it eventually. The distance of the trajectory to the limit cycle does not decrease monotonically. It initially decreases, then increases and eventually decreases as shown in fig. \ref{fig:traj_limit_cycle}(b). The initial conditions chosen for this example are generic, containing perturbations from the limit cycle in both the stable and unstable direction.

As in the previous example the global stability of the limit cycle depends on the value of $\omega$ for a fixed set of $\lambda_1$ and $\lambda_2$. In this case too, there are four different regimes of $\omega$ where the behavior of trajectories is qualitatively  different.  Exploring a range of values of $\omega$, we find that for small values of $\omega$, trajectories that begin close to the limit cycle diverge with the distance to the limit cycle, $d(t)$, increasing monotonically. As the value of $\omega$ increases, this divergence is such that $d(t)$ decreases at regular intervals of time, but on the average increases, as in fig. \ref{fig:stages_z}(b). At a critical value of the rate of rotation, $\omega_{cr} \approx 22.4686859$ trajectories that begin close to the limit cycle are bounded but do not converge to the limit cycle, fig. \ref{fig:traj_limit_cycle}(a). The distance of such a generic trajectory, $(r(0) = 1.1, y(0) = 0.1, \theta(0) =0)$, is shown in fig. \ref{fig:traj_limit_cycle}(b) which oscillates periodically. The trajectory itself moves through two neighborhoods around the limit cycle, one where the distance from the limit cycle increases and one where it decreases. Numerical simulations suggest that these neighborhoods of the limit cycle where distance decays are nearly independent of the initial conditions.

\section{Conclusion}

We have shown that the invariant manifold of a dynamical system in $\mathbb{R}^n$, $n=3$ can be such that it is locally unstable everywhere, but still act as a global attractor. These examples can be trivially extended to higher dimensions, $n>3$, by increasing the dimension of the invariant manifold. It turns out that a simplified form of the Maxey-Riley equation for the motion of inertial particles in a fluid is in fact a non trivial extension of the examples considered here to $\mathbb{R}^4$, \cite{sspt_cnsns_2017}. The findings in this paper clearly attribute the cause of the global stability of the invariant manifold, $M$, to the rate of rotation, $\omega$, of the stable and unstable eigenvectors spanning the normal subspace at each point on $M$.  Our findings in this paper should be of interest from both a theoretical perspective and from the point of view of applications such as those in micro scale fluid flows where globally stable invariant regions of a fluid are sought to be created that act as particle traps.

We explored the parametric dependence of the stability of $M$ on $\omega$ in this paper for fixed values of the rates of expansion and contraction in the normal direction. It should be obvious that choosing a different set of values of $\lambda_1$ and $\lambda_2$ while preserving the relationship $\lambda_1 + \lambda_2 <0$, will yield different critical values of $\omega$. Furthermore in the two examples considered here, the dynamics on the invariant manifold are slow, due to the constant multiplicative factor of $0.01$ in \eqref{eq:sys} and \eqref{eq:sys2}. Choosing a different multiplicative factor changes the values of $\omega$ at which transitions between the stable and unstable regimes occur. When this multiplicative factor is large, $1$, comparable to the $|\lambda_1|$, the limit cycle is stable for very small values of $\omega <1$. We do not discuss these numerics in this paper, but merely point out the possibly complex dependence of stability of $M$ on a large set of parameters. 

%\section{Acknowledgment}
%We thank the Reviewers for their comments and raising important questions that led to a significant  improvement in the quality of the presentation of results.

%\bibliographystyle{unsrtnat}
%\bibliography{bibliography}

\end{document}